\pdfoutput=1
\documentclass[12pt,
				a4paper,
				twoside,
				appendixprefix,
				abstracton,
				BCOR8mm]{scrreprt}  

\usepackage[utf8]{inputenc}
\usepackage{amsmath}
\usepackage{amssymb}
\usepackage{graphicx}
\usepackage{epstopdf}
\usepackage{thumbpdf}
\usepackage{color}
\usepackage{url}
\usepackage{geometry}
\usepackage{tabularx}
\usepackage[version=3]{mhchem}
\usepackage{overpic}
\usepackage[url=false,sorting=none,firstinits=true,isbn=false,arxiv=abs,maxbibnames=99]{biblatex}

\usepackage{pxfonts}

\usepackage{user}

\graphicspath{{img/}}
\AtEveryBibitem{
  \clearfield{day}
  \clearfield{month}
  \clearfield{endday}
  \clearfield{endmonth}
  \clearfield{issue}
  \clearfield{number}
}

\DeclareFieldFormat{postnote}{\mkpageprefix[pagination]{#1}}
\DeclareFieldFormat{pages}{\mkpageprefix[pagination]{#1}}

\begin{document}


\newcommand{\dcsubject}{Bachelor Thesis}
\newcommand{\dctitle}{Reduction of Copper Oxide by Formic Acid}
\newcommand{\dcsubtitle}{an ab-initio study} 

\newcommand{\dcauthorlastname}{Schmeißer}
\newcommand{\dcauthorfirstname}{Martin}
\newcommand{\dcauthormiddlenames}{Anton Helmut}
\newcommand{\dcauthoremail}{martin.schmeisser@s2008@tu-chemnitz.de} 
\newcommand{\dcdate}{\the\day. \monthword{\the\month}~\the\year}

\newcommand{\dcplace}{Chemnitz} 
\newcommand{\dcuni}{\dcplace~University of Technology}
\newcommand{\dcunidepart}{Faculty of Natural Sciences}
\newcommand{\dcunilogo}{TUC_english_CMYK}
\newcommand{\dcinstitute}{Fraunhofer Institute for Electronic Nano Systems}
\newcommand{\dcinstdepart}{Department Back-end of Line}
\newcommand{\dcinstlogo}{FH_ENAS_Logo_60mm_CMYK}
\newcommand{\dcprof}{} 

\newcommand{\dcpruefer}{Prof. Dr. Stefan E. Schulz\\\>Fraunhofer ENAS\\\>Honorary Professorship "Technologien der Nanoelektronik"\\\>Faculty of Electrical Engineering and Information Technology\\\>Chemnitz University of Technology}
\newcommand{\dcsecpruefer}{Prof. Dr. Alexander Auer\\\>Max-Planck-Institute for Iron Research\\\>Honorary Professorship "Computational Quantum Chemistry"\\\>Faculty of Natural Sciences\\\>Chemnitz University of Technology}
\newcommand{\dcadvisor}{Dr. Jörg Schuster\\\>Fraunhofer ENAS}

\newcommand{\dckeywords}{copper oxide,CuO,formic acid,HCOOH,reduction,ab initio,density functional theory,DFT,atomic layer deposition,ALD}

\hypersetup{%
	pdftitle	= {\dctitle}, %
	pdfsubject	= {\dcsubject, \dcdate}, %
	pdfauthor	= {\dcauthorfirstname~\dcauthorlastname, \dcauthoremail}, %
	pdfkeywords	= {\dckeywords}, %
	pdfcreator	= {pdfTeX with Hyperref and Thumbpdf}, %
	pdfproducer	= {LaTeX, hyperref, thumbpdf}, %
}


\titlehead{

\parbox{6cm}{\centering \includegraphics[width=6cm]{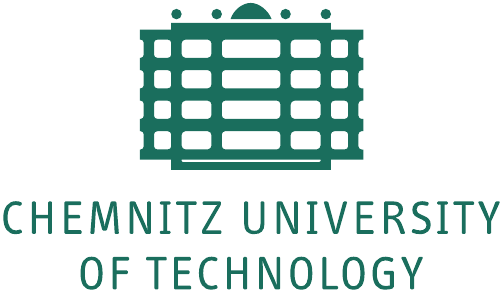} } 
\hfill
\parbox{6cm}{\centering \includegraphics[width=6cm]{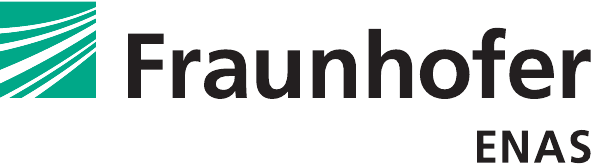} }\\
\hrulefill

}

\subject{\bf\Huge\dcsubject}

\title{\Large
	\dctitle
	\\
	\dcsubtitle
}

\author{\dcauthorfirstname~\dcauthorlastname}
	
\date{\dcplace, \dcdate\\
}

\publishers{
	{
\small
	\parbox{0pt}{
		\begin{tabbing}
			{\bf second examiner:}\quad\=\kill
			{\bf Examiner:}	\>\dcpruefer\\
			{\bf Second examiner:}	\>\dcsecpruefer\\
		\end{tabbing}	
	}}
}

\lowertitleback{
\small

\textbf{\dcauthorlastname, \dcauthorfirstname~\dcauthormiddlenames}\\
\dctitle~-~\dcsubtitle\\
\dcsubject\\
\dcuni,~\dcunidepart\\
\dcinstitute,~\dcinstdepart\\
\monthword{\month}~\number\year\\
\vspace{0.5cm}\\
This is a corrected version, the original bachelor thesis dates to 28.09.2011 and was submitted on 29.09.2011.
}

\maketitle

\cleardoubleemptypage
\pagenumbering{roman}
\tableofcontents

\clearpage
\section*{List of Acronyms}
\begin{tabbing}
\quad\quad\quad\=\quad\quad\quad\quad\quad\=\kill
\>ALD \>atomic layer deposition\\
\>CC \> coupled cluster\\
\>CI \> configuration interaction\\ 
\>DFT  \>density functional theory\\
\>DNP  \>double numerical plus polarization basis set\\
\>EF  \>eigenvector following\\
\>GGA  \>generalized gradient approximation\\
\>HF  \>Hartree-Fock\\
\>LDA  \>local density approximation\\
\>LEED \>low energy electron diffraction\\
\>LSDA  \>local spin density approximation\\
\>LST  \>linear synchronous transit\\
\>PVD  \>physical vapour deposition\\
\>QST  \>quadratic synchronous transit\\
\>SVP \>split valence plus polarization basis set\\
\>TS  \>transition state\\
\>TSV  \>through silicon via\\
\>TZVP \>triple zeta valence plus polarization basis set\\
\>XPS  \>x-ray photoelectron spectroscopy\\
\>ZPVE  \>zero point vibrational energy\\
\end{tabbing}

\clearpage

\section*{List of Symbols}
\begin{tabbing}
\quad\quad\quad\=\quad\quad\quad\quad\quad\=\kill

\>$\Delta E$  \>reaction energy  \\
\>$\Delta E_\text{a}$  \>activation energy, barrier\\
\>$E$  \>energy (eigenvalue)  \\
\>$E_\text{XC}$  \>exchange-correlation energy\\
\>$K$  \>equilibrium constant\\
\>$ \hat{H} $\>\textsc{Hamilton}-Operator\\
\>$H$  \> enthalpy\\
\>$H_\text{vib}$  \> enthalpy due to vibrations\\
\>$k$  \>reaction rate constant  \\
\>$k_\text{b}$  \>\textsc{Boltzmann}-Constant\\
\>$R$  \>ideal gas constant\\
\>$z$  \>partition function\\
\>$ \hat{T} $\>kinetic energy operator\\
\>$ \hat{V} $\>potential energy operator\\
\>$\varepsilon$  \>energy eigenvalue of a single particle \\
\>$ \rho $\>electron density\\
\>$ \varphi $\>a single particle wave function\\
\>$ \psi $\>a many particle wave function\\
\>$  $\>\\
\end{tabbing}

\cleardoublepage
\pagenumbering{arabic}
\setcounter{page}{1}

\chapter{Introduction} \label{chap:Introduction}

\begin{figure}[!h]
\includegraphics[width=.8\textwidth]{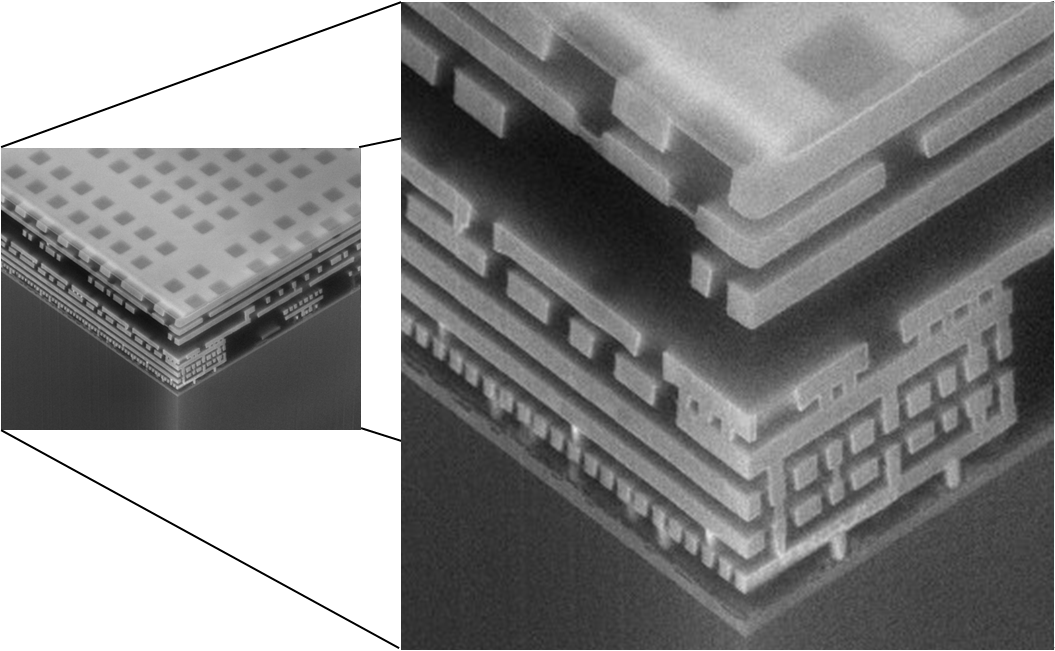}
\caption{SEM cross-section of AMD Opteron\texttrademark~and AMD Athlon\texttrademark64 microprocessors, showing the 9-metal interconnects hierarchy (IEDM 2003, Nanofair 2003), Photo courtesy AMD Saxony (now GLOBALFOUNDRIES) - Dresden, Germany.}
\label{fig:metal_SEM}
\end{figure}

\begin{figure}[!h]
\includegraphics[width=.8\textwidth]{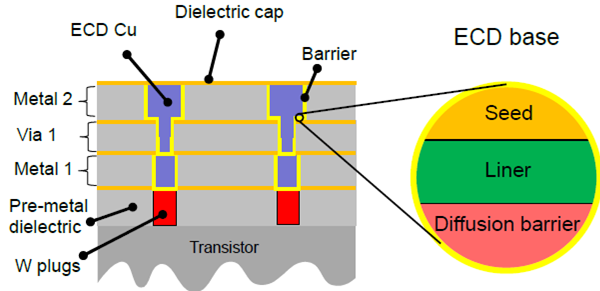}
\caption{Schematic illustration of the metallization layers in recent semiconductor devices, picture courtesy of Steve Müller.}
\label{fig:damascene}
\end{figure}

In current semiconductor technology, vias and interconnect leads are manufactured by physical vapour deposition (PVD, sputtering) of a copper seed layer and subsequent electrochemical metal deposition. Steadily decreasing feature sizes reduce the lead's diameter and thus increase the resistivity. Furthermore, current density increases, which promotes electromigration of copper ions. Also, vias with high aspect ratios are hard to fill by sputtering. Figures \ref{fig:metal_SEM} and \ref{fig:damascene} show an schematic representation of such structures.

In the strive for a sustainable metallization technology, Atomic Layer Deposition (ALD) has recently been proven as a means to grow copper oxide layers on spatially structured as well as on large plane substrates \cite{Wachtler2009}. Thomas Wächtler presented a hybrid approach to divide the metallization approach into an copper oxide ALD and a subsequent reduction step. Direct deposition of metallic copper was not possible in the desired quality because copper tends to form islands and clusters (agglomeration), but the field is also actively investigated \cite{Lee2009}.

ALD is a promising technology because it is known to produce void free and homogeneous films. Benefits are expected since it is possible to deposit a seed layer in high-aspect-ratio vias by ALD and because films can be deposited more homogeneously than with current technology. A film with less voids will exhibit lower resistivity and electromigration. The major obstacle in employing this technology is the reduction of copper oxide to metallic copper, which has to be carried out at low temperatures to avoid agglomeration of the resulting copper film.

\clearpage

The actual reduction reaction on the samples is hard to characterize experimentally because the grown film is not epitaxial and contains both copper(I)- and copper(II)-oxide. Moreover, measurements would have to be performed without exposing the samples to ambient conditions after the ALD process, and spectroscopic equipment is currently not available at the ALD chamber. In current investigations samples were exposed to air during transport between the ALD chamber and the spectroscopic measurements. Studying the most basic reactions in a theoretical work at the electronic structure level is thus a first step towards a more thorough understanding of the process.

In order to gain a deeper understanding of the reactions involved from a theoretical point of view it is necessary to find a valid model system for the copper oxide surface that can be used to investigate reaction mechanisms and test proposed reactions for their energetic profile. Hence, the first task in this work will be to select a feasible copper oxide structure for electronic structure calculations. Due to the accuracy requirements and the system size of about 100 atoms, density functional theory (DFT) appears as the optimal choice for all calculations as it delivers a reasonable balance between computational complexity and accuracy. This is discussed in more detail in section \ref{sect:DFT}.


Formic acid and its active decomposition products, hydrogen and carbon monoxide, are known to act as reduction agents for copper oxide. Decomposition of formic acid and reactions with different adsorbed species on the surface can be investigated once a surface model is available.

\section{Preliminary Work} \label{chap:prev}
During a previous internship at Fraunhofer ENAS the unimolecular thermal decomposition of formic acid in gas phase was modelled  \cite{my_Schmeisser2011}. Two possible decomposition paths with similar reaction barriers and enthalpies were found:

\begin{align}
HCOOH &\longrightarrow CO_2 + H_2 &\Delta E = -3.5 \, \frac{\text{kcal}}{\text{mol}}\\
HCOOH &\longrightarrow CO + H_2O  &\Delta E = 6.3 \, \frac{\text{kcal}}{\text{mol}}
\end{align}

Reaction barriers were estimated to 70 kcal mol$^{-1}$ for both reactions. Reaction energies above are from \cite{my_CRCHandbook}. Essentially the conclusion was, that under the reactor conditions described in Thomas Wächtlers work \cite{Wachtler2009}, formic acid will not dissociate fast enough to supply noticeable amounts of CO or \H2 to the sample. The introduction to chemical kinetics in the Eyring model is a result of that work (See section \ref{sect:Kinetics}).
 
\section{Known Reactions and Issues} \label{chap:literature}

Several experiments on reduction of copper oxide have been performed by Müller \textit{et al.} \cite{my_MuellerSCD2011}. It has been shown that reduction of a CuO film to metallic copper can be done by hydrogen radicals on a ruthenium or nickel substrate, a reduction to \Cu2O is possible by molecular hydrogen on these substrates. On a cobalt substrate reduction proceeds to \Cu2O with hydrogen radicals. All reductions have been carried out below 300$^\circ$C to avoid agglomeration of the deposited film. Complete Reduction by molecular hydrogen is possible at higher temperatures, but then agglomeration becomes an issue.

Using formic acid as a reduction agent, Wächtler and Müller observed no reaction on a tantalum nitride substrate, but complete reduction to metallic copper took place on a ruthenium substrate \cite{my_MuellerSCD2011}. Suspecting a catalytic influence of ruthenium, films were prepared with a mixed copper and ruthenium precursor on TaN. It was shown, that the films contained 1 to 5 percent of ruthenium and that all CuO fractions could be reduced by formic acid. To date, it is not clear whether remaining \Cu2O fractions are due to incomplete reduction or due to re-oxidation during transport of the samples.

From other experiments it is known, that clean Ruthenium and Nickel surfaces catalytically promote the decomposition of formic acid \cite{Columbia1994}.

Poulston \textit{et al.} performed temperature-programmed desorption studies of formic acid decomposition on both CuO and \Cu2O~\cite{Poulston1998}. They report a formate adsorption of formic acid, formate decomposition to gaseous CO$_2$ and adsorbed hydrogen as well as desorption of molecular hydrogen and water formed from adsorbed hydrogen and lattice oxygen. Formic acid was adsorbed at 300 K and two desorption peaks of all species occur at 430 K and 545 K, where the latter one is more intense.

Reduction of \Cu2O with CO and H or H$_2$ is already well understood at high temperatures because of its application in automotive exhaust catalysts.

The reduction of copper oxides with hydrogen and carbon monoxide has been investigated by Kim, Wang,~\textit{et al.} \cite{Kim2003, Wang2004}. In a temperature programmed reduction using carbon monoxide as reduction agent, the reduction of CuO was found to proceed either directly to metallic copper when high amounts of CO were supplied or via formation of \Cu2O~when CO supply was limited. In both cases a temperature dependant induction period was observed. The reduction of \Cu2O~was reported to proceed slower than the reduction of CuO.
Similarly, in a constant-temperature reduction using molecular hydrogen as a reduction agent, an induction period involving the embedding of hydrogen into the bulk oxide was reported from in-situ time-resolved x-ray diffraction. The activation barrier was, again, higher for \Cu2O (27.4 kcal mol$^{-1}$) than for CuO (14.5 kcal mol$^{-1}$). Additionally, density functional calculations were carried out on a bulk model to explain the transition from CuO to \Cu2O instead of Cu$_4$O$_3$.

Goldstein and Mitchell have recently measured reaction rates of copper oxide reduction with carbon monoxide using copper(I) and (II) oxide powder in a pressurized thermogravimetric analyser \cite{Goldstein2011}. They report an activation energy of 20 kJ mol$^{-1}$ and 25 kJ mol$^{-1}$ for CuO and \Cu2O, respectively.

\begin{figure}[!hb]
\begin{center}
\includegraphics[width=0.4\textwidth]{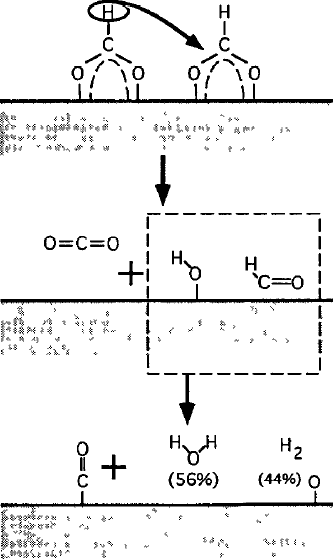}
\end{center}
\caption{Reaction model for bimolecular decomposition of formate on Ru, as proposed by Sun and Weinberg \cite{Sun1991}. Figure reprinted from \cite{Columbia1994}.}
\label{fig:columbia}
\end{figure}

On Ruthenium, a reaction model was proposed by Sun and Weinberg \cite{Sun1991} where two formate intermediates stabilize on neighbouring sites and  a "hot hydrogen" breaks the C-O bond of a second formate, resulting in one desorbed \CO2, one adsorbed hydroxyl(-OH) and one adsorbed formyl(-CHO). The adsorbed species react to form adsorbed CO and either desorbed \H2O or desorbed \H2 and adsorbed oxygen. Figure \ref{fig:columbia} illustrates the mechanism.

Copper oxide surface structures have been investigated by several groups with a variety of experimental and theoretical methods.

Schulz and Cox performed photoemission and low-energy-electron-diffraction (LEED) studies of \Cu2O (111) and (100) surfaces \cite{Schulz1991}. The (111) surface  was reported to stabilize in a stoichiometric (1x1) form after ion bombardment and annealing in vacuum. A $(\sqrt{3} \times \sqrt{3})R30^\circ$ periodicity was also observed during studies of catalytic chemistry due to a $\frac{1}{3}$ monolayer of oxygen vacancies. The (100) surface exhibited four different periodicities in the LEED studies. A reconstructed, copper terminated surface with a $(3\sqrt{2} \times \sqrt{2})R45^\circ$ periodicity  could be prepared by ion bombardment and vacuum annealing. After high oxygen exposures an unreconstructed, $(1 \times 1)$, oxygen terminated surface was reported.

\begin{figure}[!hb]
\begin{center}
\includegraphics[width=\textwidth]{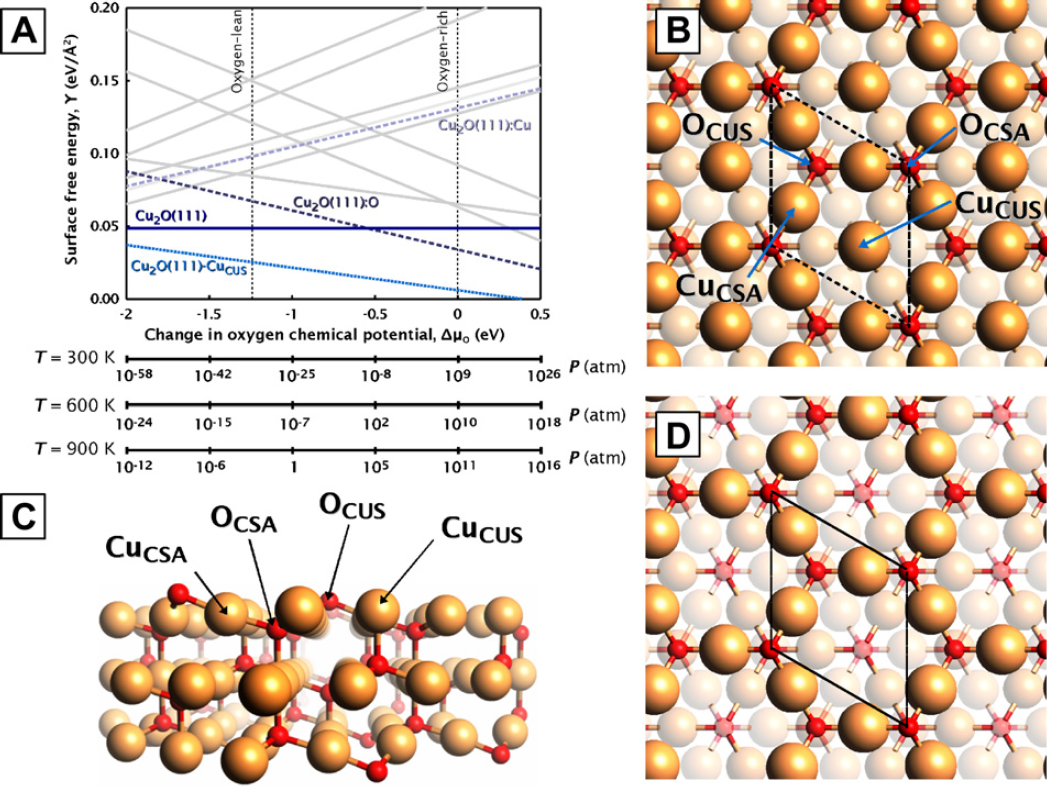}
\end{center}
\caption{\textbf{A} plots the surface free energy for the various \Cu2O(111) surface terminations as a function of oxygen chemical potential. \textbf{B} and \textbf{C} show the top and side view of the optimized atomic structure of Cu2O(111), respectively. \textbf{D} shows the top view of the low-energy structure without \CUS~species. \newline
The upper Cu atoms are shown as large orange spheres, with the oxygen atoms denoted by small red spheres.
The labels "CSA" and "CUS" stand for coordinatively saturated, and coordinatively unsaturated, respectively. Note that O$_\text{CUS}$ and O$_\text{CSA}$ will be called O$_\text{SUF}$ and O$_\text{SUB}$ in this work, respectively. Figure reprinted from \cite{Soon2007}.}
\label{fig:soon_surface}
\end{figure}

A DFT study by Soon \textit{et al.} predicted a Gibbs free energy preference for a Cu-lean \Cu2O (111) surface where the \CUS~species are vacant above 300 K and under oxygen exposure \cite{Soon2007}. Figure \ref{fig:soon_surface} shows the calculated free energies (a), an optimized bulk terminated surface structure (b and c) and the low energy structure with \CUS~vacancies (d).

Another experimental study by Önsten \textit{et al.} using LEED and scanning tunnelling microscopy (STM) resulted in two models for the (111) surface \cite{Oensten2009}. In the first model the $(1 \times 1)$ form is the ideal bulk terminated surface and the $(\sqrt{3} \times \sqrt{3})R30^\circ$ form is due to oxygen vacancies, in agreement with the results from Schulz and Cox. The second model agrees with the first one, but \CUS~species are vacant in both surface forms, which is supported by the findings of Soon \textit{et al}.

The polar \Cu2O (100) Cu$^+$ terminated surface was investigated by Nygren \textit{et al.} in a theoretical model \cite{Nygren1996a}. The $(1 \times 1)$ reconstruction with half of the Cu$^+$ placed at the opposite site of the crystal gave the lowest surface energy. The $(3\sqrt{2} \times \sqrt{2})R45^\circ$ reconstruction was not stable within their model.

\clearpage

\section{Overview of Reactions and Species involved in Formic Acid Decomposition} \label{chap:reactions}
In the following section a list of reactions is presented that an formic acid molecule might undergo. This list is not exhaustive, other reactions may occur, but the most likely reactions are listed and may be used as a starting point for later investigations once a copper oxide surface model is available.

\vspace{0.5cm}

Formic acid may decompose thermally prior to adsorption :
\begin{align}
\cee{
HCOOH &-> CO2 + H2\\
HCOOH &-> CO + H2O
}
\end{align}
or adsorb and then proceed to possible decomposition products:
\begin{align}
\cee{
HCOOH_{(g)} &-> HCOOH_{(ads)}\\
HCOOH_{(ads)} &-> H_{(ads)} + HCOO_{(ads)}\label{rctn:HCOOH_acid_decomp}\\
	&-> OH_{(ads)} + HCO_{(ads)}\\
	&-> CO_{(ads)} + H2O_{(ads)}\label{rctn:HCOOH_CO_decomp}
	}
\end{align}
The reaction mechanism for formic acid decomposition on Ru proposed by Sun and Weinberg may as well be existent on Cu$_\text{x}$O
\begin{align}
\cee{
2(HCOO_{(ads)}) &-> CO2_{(g)} + OH_{(ads)} + HCO_{(ads)} \\
OH_{(ads)} + HCO_{(ads)} &-> CO_{(ads)} + H2O_{(g)}\\
						 &-> CO_{(ads)} + O_{(ads)} + H2_{(g)}
	}
\end{align}
Different reactions with previously adsorbed species are possible
\begin{align}
\cee{
	HCOOH_{(ads)} + H_{(ads)}	&-> HCOO_{(ads)} + H_{2(g)}\label{rctn:acidH+adsH}\\
								&-> HCO_{(ads)} + H2O_{(ads)}\\
	HCOOH_{(ads)} + OH_{(ads)}	&-> HCOO_{(ads)} + H2O_{(ads)}\label{rctn:acidH+adsOH}\\
								&-> COOH_{(ads)} + H2O_{(ads)}\\
    }
\end{align}
as well as reactions with surface oxygen and adsorbed species
\begin{align}
\cee{
	O_{SUF}-HOOCH +  H_{(ads)} &-> HCOO_{(ads)} + H2O_{(g)} \label{rctn:OSUF+adsH}\\
    }
\end{align}

A few of these reactions will be modelled to test the model system for its applicability.
\chapter{Theoretical Background} \label{chap:Theory}

Electronic structures are evaluated by means of density functional theory, which is considered the standard method for simulations at this scale because it allows easy (in terms of computer time) treatment of systems of up to some hundreds of atoms at reasonable accuracy. Other, wave function based, techniques are available, where MP2 has comparable accuracy and is applicable to systems of 50 to 100 atoms. High accuracy methods like CCSD(T) and CI reach their limits at about 20 atoms. In order to predict possible reaction paths and to estimate reaction rates, reactants and products of a reaction are modelled at atomic scale and structures are optimized. Structure optimization means to find the atomic coordinates that exhibit a locally minimal DFT energy $E_\text{0}$. Comparing the reactants' and products' energy sums yields the reaction energy $\Delta E$.

\begin{equation}
\Delta E = \sum\limits_\text{products}E_\text{0,p} - \sum\limits_\text{reactants}E_\text{0,r}
\end{equation}

Vibrational effects are accounted for in a harmonic oscillator model, yielding respective enthalpies. $H_0$ is the enthalpy that includes only the zero point vibrational energy (ZPVE), when higher vibrational states become occupied the enthalpy will be a function of temperature $ H(T) $.

The reaction enthalpy will only provide insight as to whether the reaction is endothermic or exothermic. Estimating reaction rates requires knowledge of the activation energy. It is defined by the structure (and thus relative energy ) of the reaction's intermediate structure (transition state, TS). A transition state is characterized by a saddle point on the energy hypersurface.

\begin{figure}[!hb]
\includegraphics[width=\textwidth]{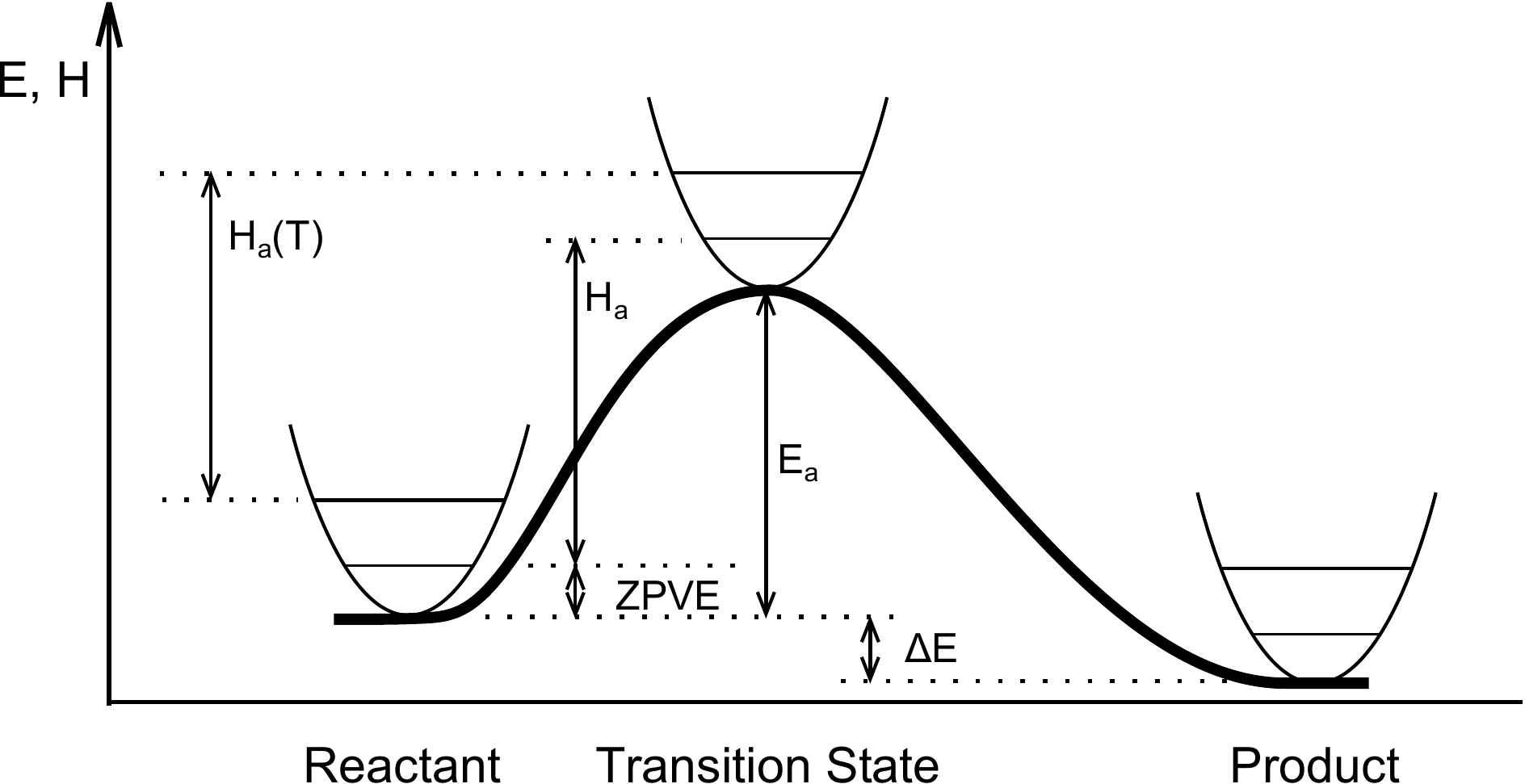}
\caption{Energy profile of a hypothetical reaction, containing definitions of the various energy and enthalpy terms. Parabolas illustrate an harmonic approximation of the potential around the structure with discrete vibrational states.}
\label{fig:enthalpy_terms}
\end{figure}

Throughout this work, $\Delta E$ and $\Delta H$ will be reaction energies and enthalpies, $E_\text{a}$ will be an activation energy. $H_\text{a}$ denotes an activation enthalpy evaluated between the zero point vibrational energies and $H_\text{a}(T)$ the activation enthalpy at finite temperature. See Figure \ref{fig:enthalpy_terms} for an illustration of the various energy and enthalpy terms.

The following chapter shall first introduce the path from a quantum mechanical many body problem to the basic DFT formalism. Then, this technique is applied to geometry optimizations, transition state searches and the transition state model for chemical kinetics.

\clearpage

\section{The Schrödinger-Equation} \label{sect:QMproblem}
In quantum theory, the fundamental problem is to find the many-body wave function $\psi$ of the system. One can then evaluate expectation values of physical properties by applying the corresponding operator to the wave function. To find the wave function, the energy eigenvalue problem (\textsc{Schrödinger}-Equation) must be solved. \textsc{Schrödinger}'s-Equation in time-independent form is:
\begin{equation}
\hat{H} | \psi \rangle = E | \psi \rangle
\end{equation}
where $\psi$ is the wave function and $\hat{H}$ is the \textsc{Hamilton}-operator.
The Hamilton Operator is the total energy operator and for this problem will be the sum of a kinetic energy operator $\hat{T}$ of electrons and nuclei and potential energy operators $\hat{V}$ that describe the electron-nucleon interaction and the mutual interaction among either electrons or nuclei. In atomic units (a.u.) these are:
\begin{align}
\hat{H} = \, &\hat{T} + \hat{V}\\
\hat{H} = \, &\hat{T}_\text{e} + \hat{T}_\text{n} + \hat{V}_\text{ne} +\hat{V}_\text{ee} +\hat{V}_\text{nn}\\
 =\nonumber &- \sum\limits_{i=1}^N\frac{1}{2}\nabla^2_\text{i} - \sum\limits_{A=1}^M\frac{1}{2M_A}\nabla^2_\text{A} - \sum\limits_{i=1}^N\sum\limits_{A=1}^M\frac{Z_\text{A}}{r_\text{iA}}\\
 &+ \sum\limits_{i=1}^N\sum\limits_{j>i}^N\frac{1}{r_\text{ij}} + \sum\limits_{A=1}^M\sum\limits_{B>A}^M\frac{Z_\text{A} Z_\text{B}}{r_\text{AB}}
\end{align}
For electrons and nuclei indices ($i,A$) and position vectors ($\vec{r_\text{i}}, \vec{R_\text{A}}$) are in lower and upper case, respectively. The first two terms cover the kinetic energy of electrons and nuclei, where $\nabla^2_i$ and $\nabla^2_A$ are the Laplacian differentiation operators with respect to the coordinates of the $i$th electron and $A$th nucleus. Z$_\text{A}$ is the nuclear charge of atom A. The latter three terms represent the coulomb interaction between electrons and nuclei, electron pairs and nucleus pairs where $r_\text{ij}, r_\text{iA}, r_\text{AB}$ are the mutual distances between two electrons, one electron and one nucleus and between two nuclei.

One common simplification is the \textsc{Born-Oppenheimer}-Approximation \cite{Born1927}: the set of spatial variables is divided into electronic coordinates $\vec{r}$ and nuclear coordinates $\vec{R}$ and the wave function is expressed as the product of two separate ones for electrons and nuclei
\begin{equation}
\psi(\vec{r}) = \psi_\text{e}(\vec{r},\vec{R}) \cdot \psi_\text{n}(\vec{R})
\end{equation}
Applying this separation to the Schrödinger-Equation results in equations for electrons
\begin{align}
\hat H\cdot\psi_\text{e}(\vec{r}, \vec{R}) &= E_\text{e}\cdot\psi_\text{e}(\vec{r}, \vec{R})\\
(\hat{T}_\text{e} + \hat{T}_\text{n} + \hat{V}_\text{ne} +\hat{V}_\text{ee})\cdot\psi_{e}(\vec{r}, \vec{R}) &= E_\text{e}\cdot\psi_\text{e}(\vec{r}, \vec{R})
\end{align}
and nuclei
\begin{align}
\hat H\cdot\psi_\text{n}(\vec{R}) &= E_\text{n}\cdot\psi_\text{n}(\vec{R})\\
(\hat{T}_\text{n} +\hat{V}_\text{nn})\cdot\psi_\text{n}(\vec{R}) &= E_\text{n}\cdot\psi_\text{n}(\vec{R})
\end{align}
which are coupled via the $\hat{T}_\text{n}\cdot\psi_\text{e}(\vec{r}, \vec{R})$ term. Decoupling the equations by eliminating this term is known as the Born-Oppenheimer approximation.
The electronic wave function will then depend parametrically on the nuclear coordinates (~ie. $\psi_\text{e} = \psi_\text{e}(\vec{r}, [\vec{R}])$~), but they will not count as degrees of freedom in the system. This introduces the concept of a \textit{potential energy surface} on which local minima represent stable structures (see Section \ref{sect:geomOpt_TS}).

The \textsc{Born-Oppenheimer}-Approximation usually works very good. Jensen \cite{Jensen} argues that it introduces an error of the order of $10^{-4}$~au which is usually negligible when compared to other errors.

A more thorough and rigorous discussion can be found (among others) in \cite{Jensen, Koch2001, Dreizler1991}.

\section{Density Functional Theory} \label{sect:DFT}
Contrary to the wave function approach, density functional theory focuses on the electron density $\rho$ instead of the wave function:

\begin{equation}
\rho(\vec{r}) = |\psi(\vec{r})|^2 = \psi(\vec{r}) \cdot \psi^*(\vec{r})
\end{equation}

The total energy $E$ can then be expressed as a functional of electron density, where $E_\mathrm{Ne}$ is the energy due to nucleus-electron interaction, $T$ is the kinetic energy of the electrons, $J$ is the electronic coulomb interaction and $E_\mathrm{ncl}$ covers all non-classical effects (self-interaction correction, exchange, correlation).

\begin{align}
\nonumber E &= E[\rho]\\
 &= T[\rho] + J[\rho] + E_\text{Ne}[\rho] + E_\text{ncl}[\rho] \label{eqn:EnergyFunctionals}
\end{align}

Hohenberg and Kohn \cite{Hohenberg1964} have proven that the ground state electronic energy is given by a unique functional of the true ground state electron density and that the ground state electron density minimizes the total energy of the system. The electron density is a function only of the spatial variables, instead of all electron coordinates (spatial and spin), which drastically reduces the complexity of the problem. However, modern DFT codes, by suggestion of Kohn and Sham \cite{Kohn1965}, reintroduce these degrees of freedom by generating the density from a basis set of atomic orbitals (linear combination of atomic orbitals, LCAO).

In Kohn-Sham Theory an auxiliary \textit{non-interacting reference system} is constructed:

\begin{equation}
\hat{H_s} = -\frac{1}{2}\sum\limits_i^N \nabla^2_\text{i} + \sum\limits_i^N V_\text{s}(\vec{r_\text{i}})
\end{equation}

where $V_s$ is an effective local potential. This has the advantage that the main contribution to the real, interacting systems energy (kinetic energy) can be calculated exactly. The corresponding ground state wave function can then be expressed as a \textsc{Slater} determinant:
\begin{equation}
\Theta_\text{s} = \frac{1}{\sqrt{N!}}
\begin{vmatrix}
\varphi_1(\vec{x_1}) & \varphi_2(\vec{x_1}) & \cdots & \varphi_\text{N}(\vec{x_1}) \\
\varphi_1(\vec{x_2}) & \varphi_2(\vec{x_2}) & \cdots & \varphi_\text{N}(\vec{x_2}) \\
\vdots & \vdots &  & \vdots \\
\varphi_1(\vec{x_\text{N}}) & \varphi_2(\vec{x_\text{N}}) & \cdots & \varphi_\text{N}(\vec{x_\text{N}}) \\
\end{vmatrix}
\end{equation}
where the $\varphi_\text{i}$ are the eigenstates of the one electron \textsc{Kohn-Sham} operator $\hat{f}^{KS}$ and $V_\text{s}$ is chosen such that the resulting density equals that of the real, interacting system:
\begin{align}
\hat{f}^\text{KS} \varphi_\text{i} &= \varepsilon_\text{i} \varphi_\text{i} \\
\hat{f}^\text{KS} &=  -\frac{1}{2} \nabla^2 + V_\text{s}(\vec{r}) \\
\rho_\text{s}(\vec{r}) &= \sum\limits_i^N \sum\limits_s |\varphi_\text{i}|^2 = \rho_0(\vec{r})
\end{align}
The energy of a non-interacting system can now be calculated exactly, but only the energy of the interacting system is physically meaningful. The (interacting) electronic energy functional is divided as follows:
\begin{equation}
F[\rho] = T_\text{s}[\rho] + J[\rho] + E_\text{XC}[\rho]
\end{equation}
where $E_\text{XC}$ is the exchange-correlation energy, which, by comparison with equation \ref{eqn:EnergyFunctionals}, is defined as
\begin{equation}
E_\text{XC}[\rho] \equiv T[\rho] - T_\text{s}[\rho] + E_\text{ncl}[\rho].
\end{equation}
Again, $T$ is the true kinetic energy of an interacting system and $T_\text{s}$ is the kinetic energy in the non-interacting system, which we can calculate. All interaction effects except the classical coulomb interaction are summed up in the $E_\text{XC}$ term, of which no exact form is known yet.

Thus, to minimize the energy, the orbitals have to fulfil the following equation \cite[p. 45]{Koch2001}
\begin{equation}
\left( -\frac{1}{2} \nabla^2 + \left[ \int\! \frac{\rho(\vec{r_2})}{r_{12}} \,\mathrm{d}\vec{r_2} + V_\text{XC}(\vec{r_1}) - \sum\limits_A^M \frac{Z_\text{A}}{r_\text{1A}}\right] \right) \varphi_\text{i} = \varepsilon_\text{i} \varphi_\text{i}.
\end{equation}
In the square brackets, the first term is the electronic coulomb potential, the second is the potential that corresponds to the exchange-correlation energy $V_\text{XC} \equiv \frac{\delta E_\text{XC}}{\delta\rho}$ and the third is the nuclear coulomb potential. These are often referred to as an effective potential $V_\text{eff}$:
\begin{equation}
V_\text{eff} \equiv \int \! \frac{\rho(\vec{r_2})}{r_{12}} \,\mathrm{d}\vec{r_2} + V_\text{XC}(\vec{r_1}) - \sum\limits_A^M \frac{Z_\text{A}}{r_\text{1A}}.
\end{equation}
Up to this point, the Kohn-Sham Theory is \textit{exact}, there are no approximations, except Born-Oppenheimer, so far. If the true analytic forms of all functionals were known, the Kohn-Sham scheme would provide the exact energy. The approximation is introduced with the explicit form for the unknown functional of the exchange-correlation energy.

\section{Exchange-Correlation Functionals}
Exchange-correlation functionals are commonly divided into three groups: local (spin) density approximation (LDA), generalized gradients approximation (GGA) and hybrid-functionals. The basis for most functionals is the uniform electron gas, which is the model system for LDA, non local effects are then added in by GGA functionals.

In the model of an uniform electron gas, electrons move on a uniform positive background charge and the electron density is constant. Despite its simplicity it is a valuable model because it is the only system where the exact exchange functional is known and accurate numerical approximations to the correlation functional are available.

In the local density approximation, $E_\text{XC}$ depends only on the local electron density:
\begin{equation}\label{eqn:LDA}
E_\text{XC}^\text{LDA}= \int \! \rho(\vec{r}) \varepsilon(\rho(\vec{r})) \,\mathrm{d}\vec{r}
\end{equation}
where $\varepsilon(\rho(\vec{r}))$ is the exchange-correlation energy per particle of a uniform electron gas of density $\rho(\vec{r})$, which is weighted by the probability $\rho(\vec{r})$ of actually finding an electron at $\vec{r}$. The exchange energy of an uniform electron gas has been derived by Bloch and Dirac \cite{Dirac1929}. It is given by
\begin{equation}
\epsilon_\text{X} = -\frac{3}{4} \sqrt[3]{\frac{3 \rho(\vec{r})}{\pi}}.
\end{equation}
The correlation part has been evaluated numerically by Ceperly and Alder \cite{Ceperley1980}, and analytical expressions have been interpolated. Commonly used LDA functionals are VWN, due to Vosko, Wilk and Nusair \cite{Vosko1980} and the one due to Perdew and Wang \cite{Perdew1992a}. Extending LDA to the spin unrestricted case results in the local spin density approximation (LSDA), where
\begin{equation}\label{eqn:LSD}
E_\text{XC}^\text{LSD}= \int \! \rho(\vec{r}) \varepsilon(\rho_\alpha(\vec{r}), \rho_\beta(\vec{r})) \,\mathrm{d}\vec{r}.
\end{equation}
Note that in Equations \ref{eqn:LDA} and \ref{eqn:LSD} the non uniform density $\rho$ is inserted, but $\varepsilon$ is always from the uniform electron gas. This is a drastic approximation, but performs surprisingly well, for example it yields 36 kcal/mol average unsigned deviation on the atomization energies of the G2 test-set (compare Hartree-Fock 78 kcal/mol) \cite{Riley2007}. LDA tends to overestimate binding energies.

In the generalized gradients approximation, LDA is interpreted as the first term of a Taylor-expansion, which is continued to the next term as in
\begin{equation}
E_\text{XC}^\text{GEA} = E_\text{XC}^\text{LDA} + \int \! C_\text{XC}(\rho) \frac{\nabla\rho}{\rho^{2/3}} \,\mathrm{d}\vec{r}.
\end{equation}
which is called gradient expansion correlation. This tends to yield results worse than those obtained from LDA, because the density depletion around an electron due to exchange and correlation interactions (the exchange-correlation hole\footnote{hole functions are an essential concept of DFT but would go beyond the scope of this work}) does not fulfil its physical properties. These properties can be enforced by setting parts of the exchange hole to zero if they are non-negative and by truncating the exchange and correlation holes to the correct sum behaviour (one and zero, respectively), which is why the these functionals are then said to work within the \textit{generalized} gradient approximation (GGA). This approach performs reasonably well \cite{Curtiss1997, Cohen2000, Delley2006, Riley2007}.
Popular choices of GGA functionals are BP, with Becke's gradient correction to exchange  \cite{Becke1988a} and Perdew and Wang's gradient correction to correlation \cite{Perdew1992a}, PW91, where both exchange and correlation gradient correction are from Perdew and Wang \cite{Perdew1992a}, and PBE, due to Perdew, Burke and Ernzerhof \cite{Perdew1996c}.

Additionally, so called hybrid functionals are available, where the total energy is mixed from separate results of different levels of theory. The exchange energy is mixed from Hartree-Fock, LDA and GGA calculations and the correlation energy is calculated from an adopted GGA functional. Mixing parameters are fitted to sets of reference molecules. It has been shown, that these functionals perform exceptionally well on molecular systems (see references on GGA performance above), but they come at the price of usually twice the computer time.
\section{The Self-Consistent-Field Procedure}
Because the effective potential depends on the density - and thus on the orbitals - one has to insert a guess density, solve for the orbitals and re-insert the resulting density. This loop has to be continued until self-consistency is reached, i.e. until the change in the density matrix or resulting energy between two steps is low enough to satisfy one's convergence criteria. This is called the \textit{self consistent field} (SCF) procedure. It has to be stressed here, that convergence criteria in practical calculations are only the error bar within the applied theory. Current density functionals are expected to yield absolute energies within an error of $\pm 25-50$ kcal/mol, relative energies can be assumed to have an error of about $\pm 5$ kcal/mol.
Additional errors emerge, when not all calculation parameters are fully converged (usually, to save computer time). A prominent example is the basis set employed to generate the density. For structural exploration a basis set of single functions for core electrons and two functions for valence electrons and additional polarization functions (split valence, for example def2-SVP from the Ahlrichs set) might be sufficient. But for energetic evaluations at least a basis set of three functions per electron plus polarization functions for the valence electrons (def2-TZVP) should be used. A quadruple zeta plus valence polarization set (QZVP) is considered close to the basis set limit for DFT.

Several reviews on convergence issues \cite{Mattsson2005} and functional performance \cite{Perdew1996, Curtiss1997, Cohen2000, Delley2006, Riley2007} are available.

\section{Geometry Optimization and Transition State Searches} \label{sect:geomOpt_TS}
A structure usually has to be optimized after modelling. Optimization means to find a stable structure that represents a minimum on the potential energy surface. Such optimizations are founded either on numerical derivatives or analytically evaluated gradients (and Hessians) of the total energy with respect to the atomic coordinates. Minima are then found with a usual Newton-Rhapson or Conjugate Gradients method.

Finding the transition state of a reaction is a very similar task to a geometry optimization, because transition states are saddle points on the potential energy surface. One needs to minimize the energy but find a maximum along one direction (the direction of the - a priori unknown - \textit{reaction path}). Practically, a Newton-Rhapson like minimum search is performed, with the restriction that the energy be maximized with respect to the eigenvector that belongs to the (single) negative eigenvalue of the energy Hessian. This is called \textit{eigenvector-following}.

This technique requires a reasonable guess structure for the transition state which has to be close enough to the true TS to exhibit a negative Hessian eigenvalue with corresponding (reaction path) eigenvector. A starting point for an eigenvector-following search might be a guessed TS structure obtained from synchronous transit methods as proposed by Halgren and Lipscomp in \cite{Halgren1977}. Here, all atomic coordinates are moved synchronously from their reactant to their product position in a linear interpolation scheme. See also Sections \ref{sect:LST} and \ref{sect:EV_following}.

\section{Kinetics} \label{sect:Kinetics}
Reaction kinetics are evaluated in terms of transition state theory, which is expected to give a good approximation in the case of thermal equilibrium (fast energy exchange of a molecule with the surrounding system). A more complex theory like Rice-Ramsperger-Kassel-Marcus is not employed because the large uncertainties in reaction barriers from DFT calculations would still render the results inaccurate. It must be noted here that transition state theory gives an upper bound to the rate constants.\footnote{This section is an excerpt of previous work \cite{my_Schmeisser2011} and only included for completeness}

For illustration, consider a simple uni-molecular reaction from species A to B with one transition state, called $X^{\ddagger}$:
$$\mathrm{A} \rightleftharpoons X^{\ddagger} \rightleftharpoons \mathrm{B}$$
The equilibrium constant for the first part of the reaction $K^{\ddagger}$ (from the reactant structures to the transition state) is
$$K^{\ddagger} = \frac{c(X^{\ddagger})}{c(A)} = \frac{z^{\ddagger}}{z_{A}}e^{- E_\text{a} / R T}$$
where $c^{\ddagger}, c_{A}$ are the concentrations of the transition state (TS) and the reactant, $z^{\ddagger}, z_{A}$ their partition functions, respectively, and $\Delta E_0$ is the difference of ground state energies between reactant and transition state. R is the ideal gas constant. $z^{\ddagger}$ may now be written as
$$z^{\ddagger}=z_{\ddagger}\frac{1}{1-e^{-h\nu / k_\text{B} T}}$$
where $\nu$ is the frequency of the vibrational mode of the TS along the reaction coordinate and $z_{\ddagger}$ is the partition function without that vibration. Since the corresponding force constant is very low, the frequency will also be very low and the exponential function is expanded into a series and all but the linear term are neglected
$$z^{\ddagger}=z_{\ddagger}\frac{1}{1-e^{-h\nu / k_\text{B} T}} \approx z_{\ddagger} \frac{1}{1-(1-h\nu / k_\text{B} T)} = z_{\ddagger} \frac{k_\text{B} T}{h \nu}$$
The TS has $2\nu$ chances per unit time to dissociate ($\nu$ in each direction). Due to the very low force constant, one may assume, that every chance for dissociation will be used. However, only half of the incidents will lead to the reaction products, the others will lead back to the reactants. Thus, the rate constant is
\begin{align}
\nonumber k &= 2 \nu \cdot \frac{1}{2} \cdot K^{\ddagger} \\
\nonumber &= \nu \cdot z_{\ddagger} \cdot \frac{k_\text{B} T}{h \nu} \cdot \frac{1}{z_\text{A}} \cdot e^{-E_\text{a} / R T}\\
\nonumber &= \frac{k_\text{B} T}{h} \cdot \frac{z_{\ddagger}}{z_\text{A}} \cdot e^{-E_\text{a} / R T}\\
&= \frac{k_\text{B} T}{h} \cdot e^{\,\Delta S(T) / R} \cdot e^{-\Delta H_\text{vib}(T) / R T} \cdot e^{-E_\text{a} / R T} \label{eqn:Eyring}
\end{align}

which is known as the Eyring Equation. Here, S is the entropy, $H_\text{vib}$ the enthalpy of the system due to vibrations (including ZPVE) and $E_\text{a}$ the activation energy of the reaction. $\Delta H_\text{vib}$ is defined as
$$\Delta H_\text{vib} \equiv H_\text{a}(T) - E_{a}$$
and can be calculated by most electronic structure programs in an harmonic oscillator approximation. The energy levels of one vibration are determined by
\begin{align}
\nonumber \epsilon_\text{n} &= \hbar \omega \left(n + \frac{1}{2}\right)\\
\nonumber \omega &= \frac{1}{2\pi}\sqrt{\frac{k}{\mu}}\\
\nonumber k &= \frac{\partial^2 E}{\partial R^2}\\
\nonumber \mu &= \frac{m_1 m_2}{m_1 + m_2}
\end{align}
where $\omega$ is the vibrational frequency, $k$ is the force constant and $\mu$ is the reduced mass. 

\chapter{Computational Details}
Structure optimizations and energy evaluations were performed with the Turbomole \cite{my_Turbomole} and DMol$^3$ \cite{Delley1990,Delley2000a} programs. Within Turbomole, the b-p functional was used. It contains Slater and Dirac's LDA exchange functional \cite{Dirac1929,Slater1951} with Becke's 1988 gradient correction \cite{Becke1988a} and Vosko, Wilk and Nusair's LDA correlation functional \cite{Vosko1980} with Perdew's 1986 gradient correction \cite{Perdew1986}. Molecular orbitals were expanded in a basis set of atomic orbitals, in Turbomole a def2-SVP (\textbf{S}plit \textbf{V}alence plus \textbf{P}olarization) basis was used for structural exploration and a def2-TZVP (\textbf{T}riple \textbf{Z}eta \textbf{V}alence plus \textbf{P}olarization) basis was used for quantitative description. The multipole accelerated resolution of identity approximation for coulomb interaction (MARI-\textit{J}) \cite{Sierka2003} was employed with corresponding auxiliary basis sets \cite{Weigend2006}.

Using DMol$^3$, the BP functional was employed, which has the same exchange term as b-p in Turbomole but implements Perdew and Wang's 1992 gradient correction \cite{Perdew1992a} to the correlation term. DMol$^3$ provides numerical orbitals, of which the DNP (\textbf{D}ouble \textbf{N}umerical plus \textbf{P}olarization) set was used.

For exploration of the potential energy surface in geometry optimizations and transition state searches total energies were converged to 
$10^{-6}$~au \footnote{Atomic units (au) are hartree ($E_\text{h}$, energy, $4.3597482(26) \cdot 10^{18}$~J), bohr ($a_0$, length, $5.29177249(24) x 10^{11}$~m) and hartree per bohr (force), see \cite{IUPAC_green_book}} per SCF cycle and to $2\cdot10^{-5}$~au per geometry step,
geometries were relaxed to a maximum force of
$4\cdot10^{-3}$~au
and a maximum displacement per iteration of $5\cdot10^{-3}$~au.

In order to evaluate the vibrational modes, the structures were re-optimized to converge with a maximum force of 
$1\cdot10^{-4}$~au
and a maximum displacement per iteration of $1\cdot10^{-4}$~au 
with a refined total energy convergence of $10^{-9}$~au and $10^{-6}$~au per SCF and geometry step, respectively. 
In order to calculate vibrations, numeric derivatives of the total energy with respect to atomic coordinates of the adsorbate were calculated by means of finite differences. Each atom was displaced by $0.05$~\AA. The def2-TZVP basis set was employed.

A full geometry optimization of the smaller cluster structures with DMol$^3$ takes 24-48 hours on 16 Opteron 8350 CPUs with shared memory, depending on the quality of the initial structure. For Turbomole, the timing is about 72 hours for both geometry optimization and evaluation of vibrational modes on 16 Opteron 2218 CPUs with distributed memory.

To find a transition state, three major algorithms are commonly in use: eigenvector following, synchronous transit and nudged-elastic-band. Synchronous transit and eigenvector following methods have been applied in this work and shall be introduced briefly.

\section{Synchronous Transit Schemes} \label{sect:LST}
The potential energy surface for a reaction may be coarsely screened by linear synchronous transit methods (LST).
Reactant and product structures are guessed and relaxed.
An LST algorithm then finds the maximum energy structure in a linear interpolation of the coordinates between reactant and product.
Coordinates are moved synchronously (with the same interpolation parameter). An estimation for the Transition State can be found in a series of Conjugate Gradient (CG) and synchronous transit steps where the minimum in the CG step is used as one image in a subsequent synchronous transit step with quadratic interpolation between the three points. This is referred to as quadratic synchronous transit (QST). Coincidence of the CG minimum and the QST maximum is then a convergence criteria for the transition state search, because they provide lower and upper bounds to the transitions state energy.

Obviously, within this theory, one has to supply \textit{a priori} knowledge or chemical intuition to guess possible reactions and model reaction products.
In a sense, one forces the system to do a reaction, and asks how probable it will be in nature.

\section{Transition State Searches using Eigenvector Following} \label{sect:EV_following}

Another technique to find possible reaction paths of a system is to evaluate vibrational normal modes. Each mode has an associated displacement vector in the orthogonalized coordinate system of the vibrations, its eigenvector.
Displacing the molecular geometry along this vector yields configurations that the system will eventually occupy when it vibrates at a finite temperature.
One approach to find reaction paths is thus to identify soft vibrations that might lead to a reaction and calculate the system energy along the corresponding eigenvectors (\textit{line scanning}).
When the energy curve along the displacement coordinate has a negative curvature, then another vibrational analysis of the distorted structure will exhibit at least one imaginary mode. One can then follow this mode to the transition state in a constrained geometry optimization as explained in \ref{sect:geomOpt_TS}.

This way, an \textit{ab initio} scan of the potential energy surface in the vicinity of a local minimum can be carried out. The most probable reactions that the system will do are investigated and the energetic evaluation of the transition states is a by-product.

Thus, theoretically, no \textit{a priori} knowledge or chemical intuition is needed to study possible reactions. However, in practice, it takes significant effort and human interaction to select meaningful vibrational modes and drive them to the actual transition state that leads to a reaction.
In addition, this method becomes problematic when angular vibrations are expressed in Cartesian coordinates and need to be deflected far from the equilibrium structure because atoms will be moved on the tangent to the bond angle instead of the real vibrations' coordinate.
\chapter{Model System}
Wang et al showed that Copper(II)oxide (CuO) reduces to metallic copper via the intermediate formation of Copper(I)oxide (\Cu2O) when a limited supply of carbon monoxide is used as reduction agent \cite{Wang2004}. They also mention that the reduction of \Cu2O to Cu is much harder than the reduction of CuO to \Cu2O. The X-ray photoelectron spectroscopy (XPS) data of the ALD-samples from Wächtler et al. exhibits a much stronger Cu(I) peak than a Cu(II) one\cite{Wachtler2009}. Therefore, and because it is the rate limiting step of the total reduction, \Cu2O has been chosen as a model system and all reactions are modelled on \Cu2O surfaces.

Copper(I)oxide has a cubic structure, that was described by Hafner and Nagel \cite{Hafner1983} and Kirfel and Eichhorn \cite{Kirfel1990}. The (111) and (100) surfaces exist naturally, but the (100) surface is polar and literature is not consistent on the surface reconstruction \cite{Schulz1991, Nygren1996a}.

\begin{figure}[!h]
	\begin{overpic}[width=\textwidth,unit=1mm]{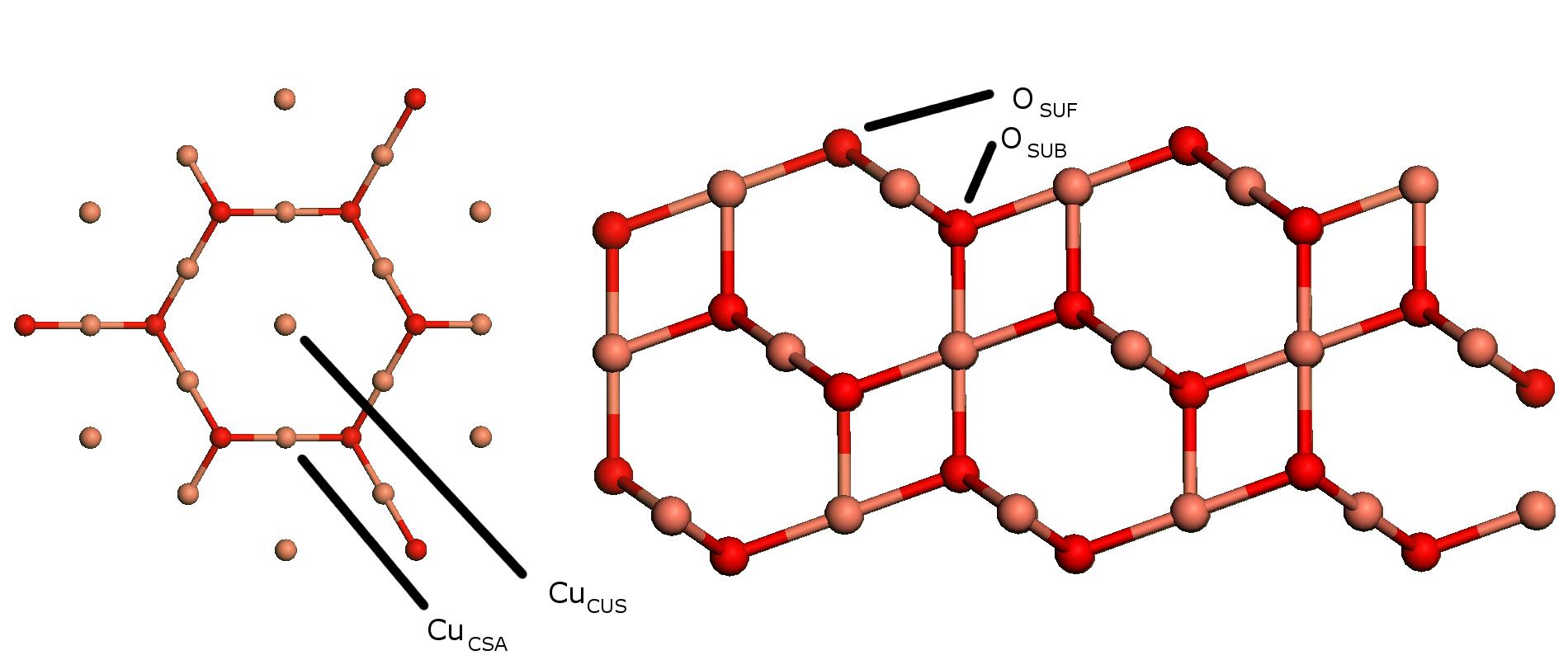}
			\put(0,40){\textbf{a)} $\bigodot(111)$}
			\put(45,40){\textbf{b)} $\downarrow(111)$}
	\end{overpic}
\label{fig:topandside}
\caption{Top (a) and side (b) view of the \Cu2O(111) surface. a) shows only one atom layer, for clarity; b) shows three atom layers as in the cluster structures that were employed. Red atoms are oxygen, orange ones are copper.}
\end{figure}

\begin{figure}[!h]
	\begin{overpic}[width=\textwidth,unit=1mm]{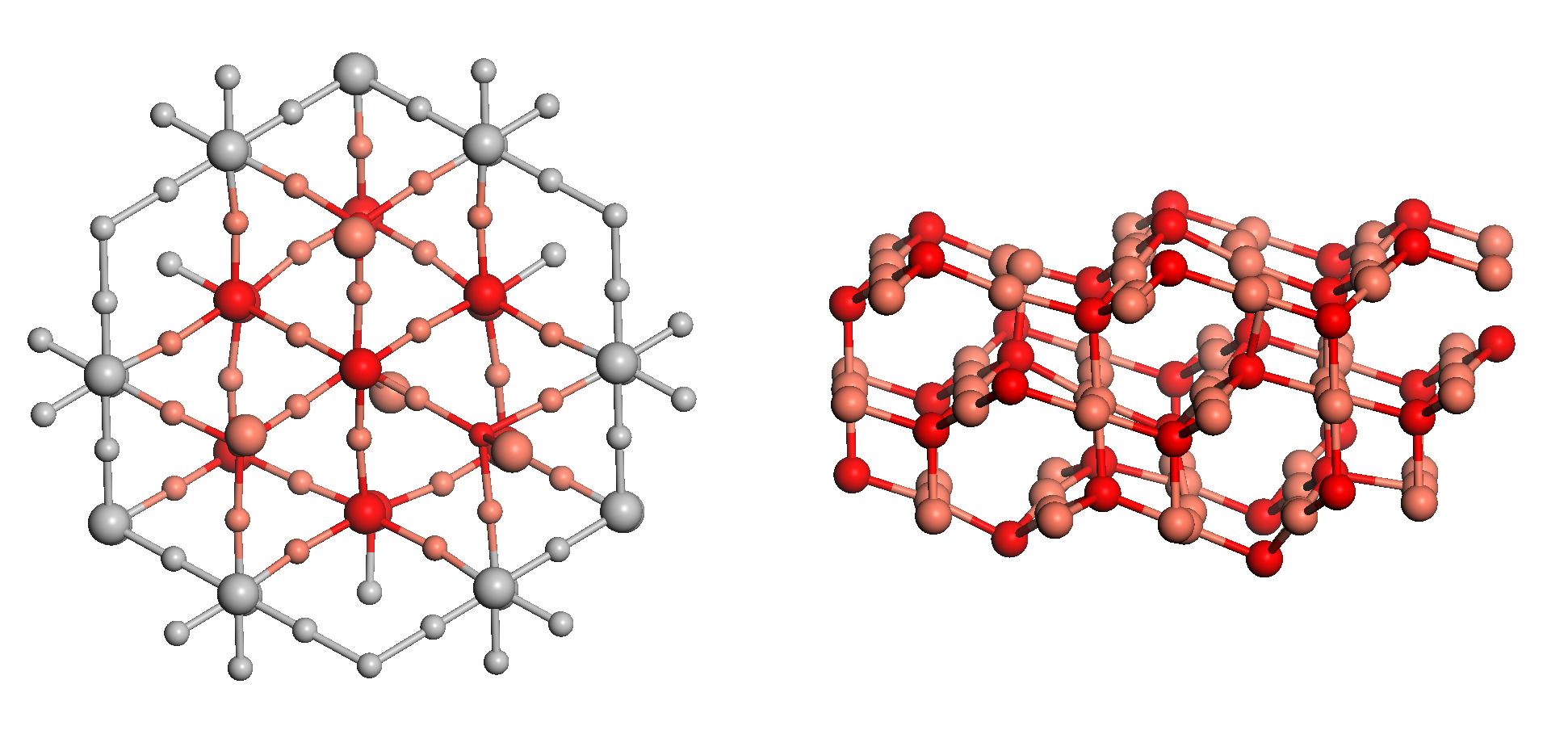}
			\put(0,40){\textbf{a)}}
			\put(45,40){\textbf{b)}}
	\end{overpic}
\label{fig:klClV1_ab}
\caption{Top (a) and side (b) view of the first Cu$_{70}$O$_{35}$ cluster (cluster 1),\newline relaxed structure, grey atoms in a) were fixed during the geometry relaxation.}
\end{figure}

\begin{figure}[!h]
	\begin{overpic}[width=\textwidth,unit=1mm]{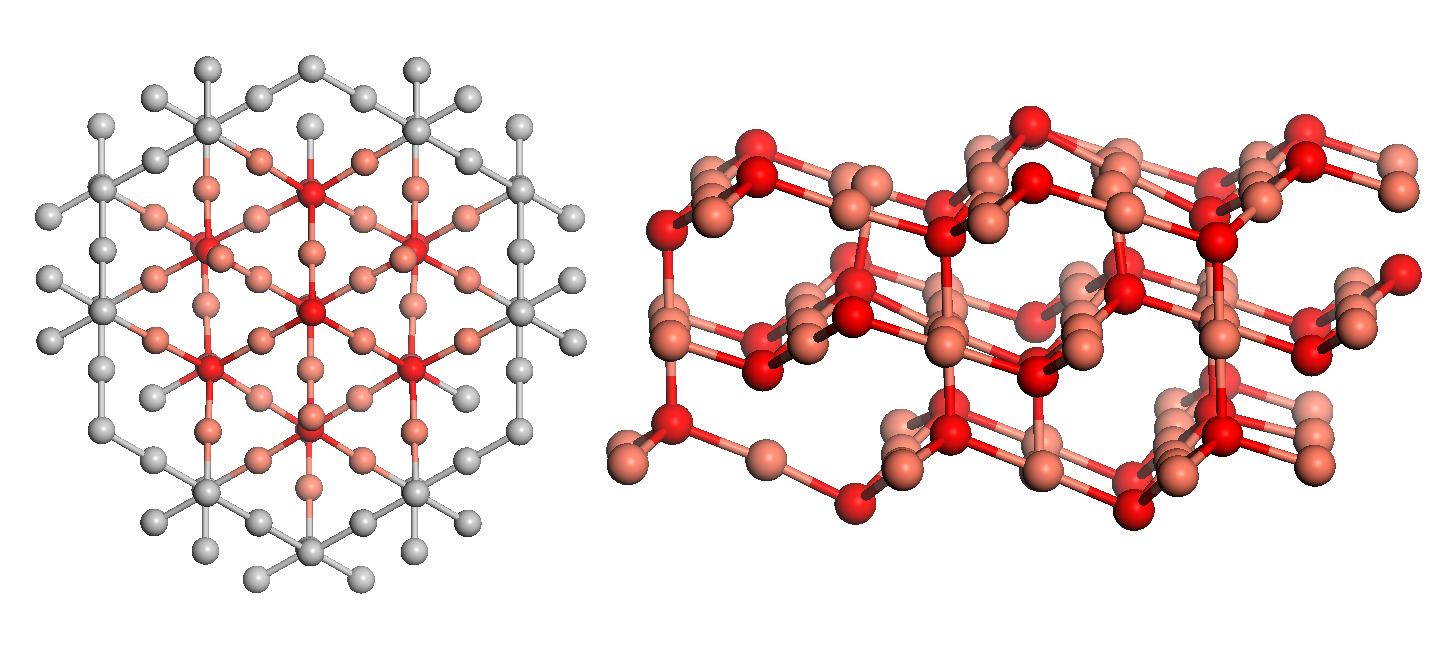}
			\put(0,40){\textbf{a)}}
			\put(45,40){\textbf{b)}}
	\end{overpic}
\label{fig:klClV2_ab}
\caption{Top (a) and side (b) view of the second Cu$_{70}$O$_{35}$ cluster (cluster 2), relaxed structure, grey atoms in a) were fixed during the geometry relaxation.}
\end{figure}

\begin{figure}[!h]
	\begin{overpic}[width=\textwidth,unit=1mm]{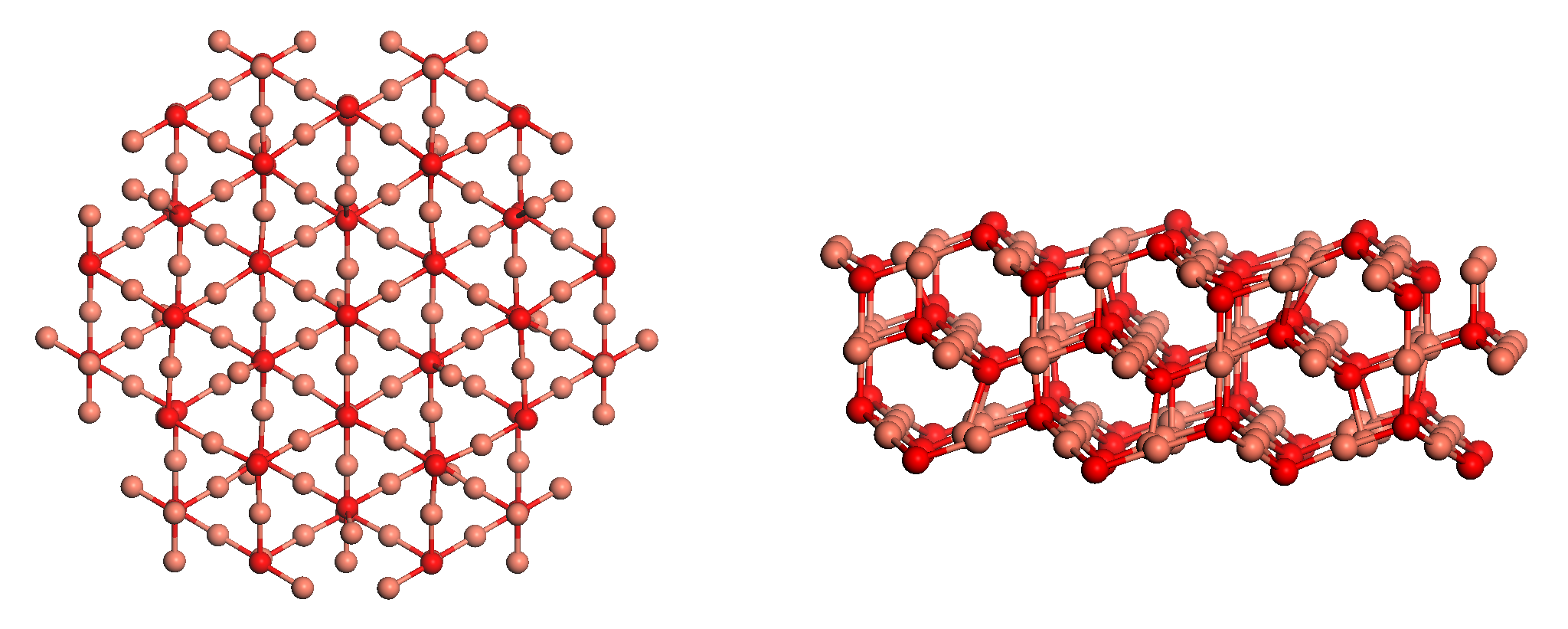}
			\put(0,35){\textbf{a)}}
			\put(45,35){\textbf{b)}}
	\end{overpic}
\label{fig:grClV1_ab}
\caption{Top (a) and side (b) view of the Cu$_{124}$O$_{62}$ cluster, relaxed structure.}
\end{figure}

Thus, for modelling, different \Cu2O(111) clusters were used. See figure \ref{fig:topandside} for the \Cu2O(111) structure. The \Cu2O(111) surface consists of copper-layers, which are embedded in two slightly shifted oxygen layers. In this work these three layers together are considered one "atom layer", contrary to e.g. \cite{Zhang2010}. On the surface, coordinatively unsaturated copper atoms exist, these are the ones with no visible bonds in figure \ref{fig:topandside}.1a, they have one coordinative bond to the underlying oxygen layer and are surrounded by a hexagon of saturated copper atoms, three oxygen atoms in a slightly raised layer and three oxygen atoms in a slightly depressed layer. Coordinatively unsaturated copper atoms are called \CUS, oxygen in the raised layer \Osuf~and oxygen in the lower layer \Osub. All clusters were stoichiometric and 3 atom layers deep, two with total formulas Cu$_{70}$O$_{35}$ (see figure \ref{fig:klClV1_ab}.2 and \ref{fig:klClV2_ab}.3), one Cu$_{112}$O$_{56}$ (not shown) and  one Cu$_{124}$O$_{62}$ (see figure \ref{fig:grClV1_ab}.4) were modelled. The two smallest clusters differ in the arrangement of the outer copper atoms.
Common to all clusters are a \CUS~species on one side and a central \Osuf~species on the other. These are the expected active sites on a clean surface, it is therefore important that they are located centrally to minimize border effects. All \Cu2O surface structures discussed in literature can easily be prepared using the existing cluster structures by removing the respective \CUS~and \Osuf~species.

\clearpage

\begin{figure}[!h]
\label{fig:slab}
\begin{center}
	\includegraphics[width=0.4\textwidth]{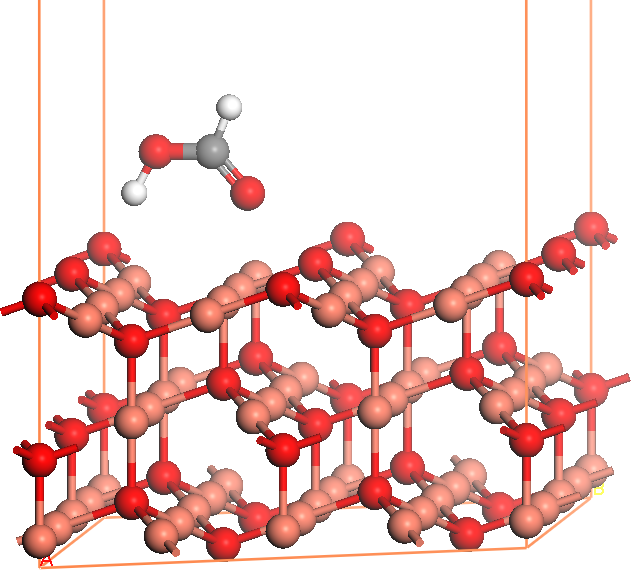}
	\caption{A slab model for the \Cu2O surface with adsorbed formic acid, \mbox{vacuum} space is cut off for better scaling.}
\end{center}
\end{figure}

Another option for a model system would be a \textit{periodic slab}. In this model a unit cell with a few atom layers of the surface is created and periodic boundary conditions are applied. A thick vacuum space is left between the surface atoms and the top face of the unit cell, see Figure \ref{fig:slab}.5 for an illustration. The advantage is, that no border effects influence the surface atoms because they reside in the potential of the infinite repetition of the unit cell. However, sufficiently large unit cells have to be used in order to minimize interactions between adsorbed species in directions parallel to the surface and interactions between periodic surface images in directions perpendicular to the surface. Such systems are best described by a plane wave ansatz for the wave function or density in a Fourier transformed space. Thus, empty space has to be treated with the same amount of basis functions, which is computationally expensive, furthermore only few density functional programs are capable of calculating modern hybrid functionals from quantum chemistry using a plane wave basis.

In comparison, the cluster model allows the application of fast quantum chemistry methods with atom centred basis sets and the full variety of available functionals. Computational effort is usually lower than with periodic systems. Energies of reactions on the surface are free of interaction with periodic images of the system but are sensitive to border effects introduced with the finite surface representation. Thus again, a sufficiently large model has to be created in order to minimize such effects. A cluster model corresponds to a low surface coverage of the adsorbed species because one adsorbed molecule is subject to the investigation while a cluster model represents a high surface coverage, depending on the size of the unit cell, an adsorbed molecule may already cover all active sites and will always interact with neighbouring cells.

\clearpage
\chapter{Results and Discussion}

\section{Geometry of the \Cu2O cluster structures}

A geometry optimization will give a first hint, whether a model is reasonable. If the structure does not distort heavily and atoms stay close to their bulk (or surface reconstructed) positions, it is likely that the model represents a valid reproduction of natural surfaces.
In a free geometry relaxation of the cluster structure, outermost copper and oxygen atoms were displaced by more than 0.9~\AA~from their bulk positions in directions parallel to the surface. Inner atoms moved only slightly, especially the hexagon ring around the central \CUS~atom was stable. In consequence, to avoid border effects and put inner atoms into a more natural electrostatic embedding, shell atoms were kept fix at their bulk positions in all subsequent calculations (see figure \ref{fig:klClV1_ab}.2a, fixed atoms are greyed out). \CUS~species moved off their bulk positions slightly when no symmetry point group was enforced, which is considered due to a numerical error.

Only the two smallest of the cluster structures were used throughout this work. Relaxed structures of the larger clusters are available, but no reactions on their surface were modelled.

Also several slab models were created with a bulk terminated surface of three atom layers. No significant reconstruction took place during geometry relaxations of the topmost layer. Again, all terminations discussed in literature may easily be created, starting from this model. These models may be used for further validation in ongoing studies.

\section{Adsorption of formic acid}

\begin{figure}[!hb]
	\centering
	\begin{minipage}[t]{.45\textwidth}
		\centering
		\begin{overpic}[width=.9\textwidth,unit=1mm]{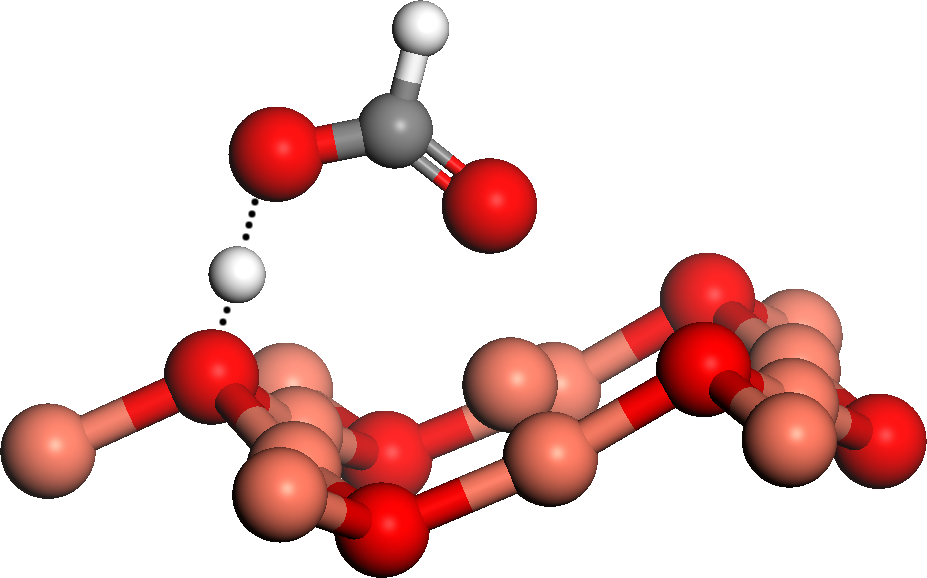}
			\put(0,70){\textbf{A}}
		\end{overpic}
	\end{minipage}
	\hspace{.05\textwidth}
	\begin{minipage}[t]{.45\textwidth}
		\centering
		\begin{overpic}[width=.9\textwidth,unit=1mm]{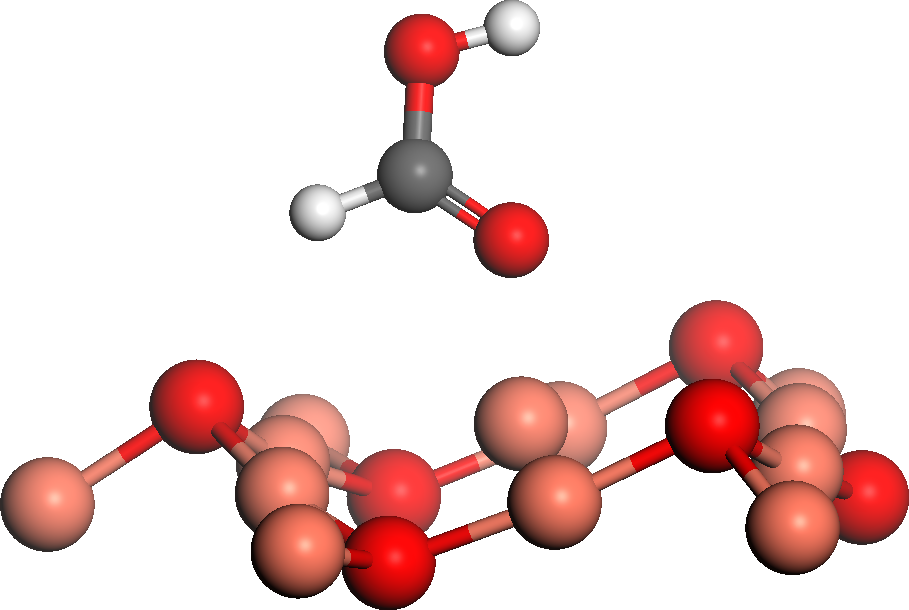}
			\put(0,70){\textbf{B}}
		\end{overpic}
	\end{minipage}

	\vspace{1cm}

	\begin{minipage}[t]{.45\textwidth}
		\centering
		\begin{overpic}[width=.9\textwidth,unit=1mm]{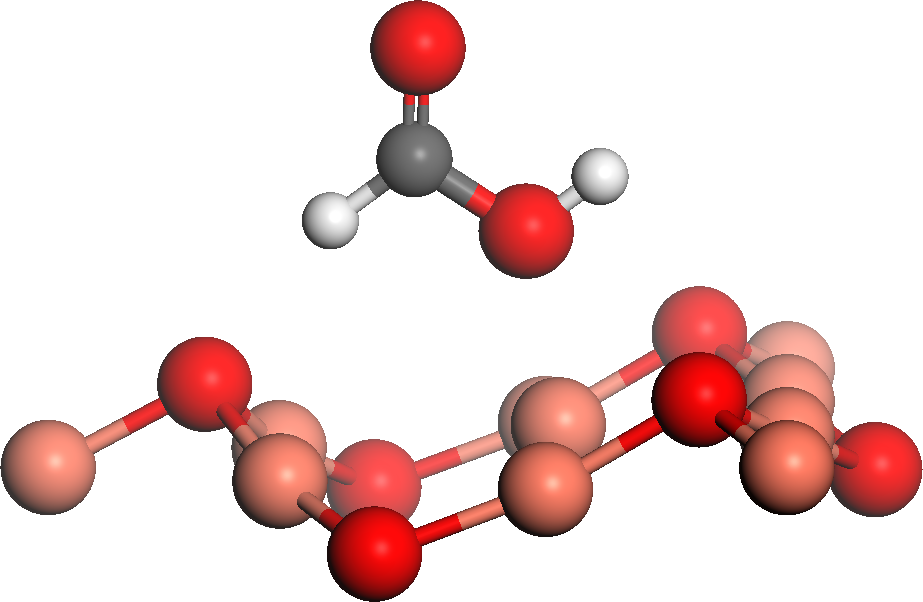}
			\put(0,70){\textbf{C}}
		\end{overpic}
	\end{minipage}
	\hspace{.05\textwidth}
	\begin{minipage}[t]{.45\textwidth}
		\centering
		\begin{overpic}[width=.9\textwidth,unit=1mm]{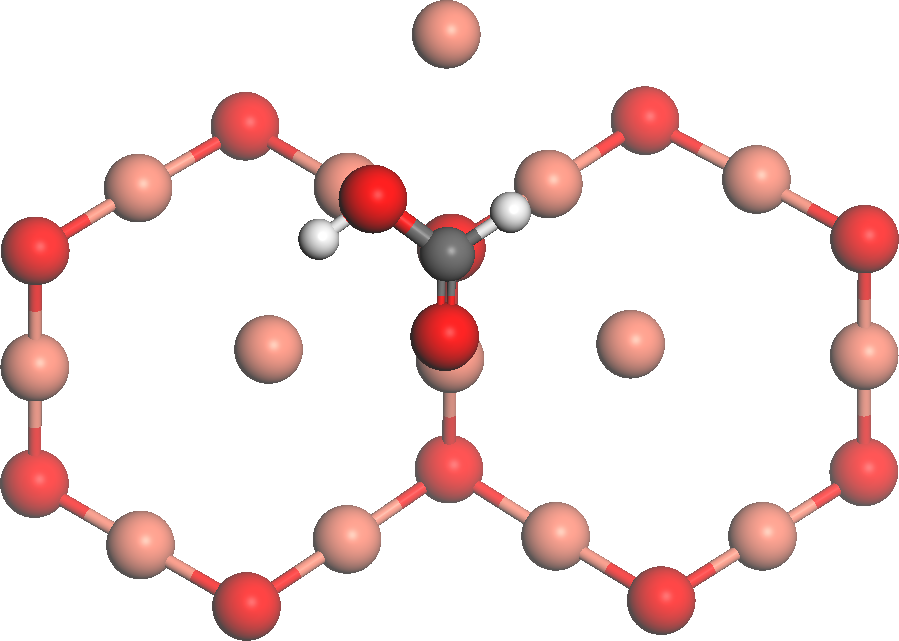}
			\put(0,70){\textbf{D}}
		\end{overpic}
	\end{minipage}
	
	\caption{Adsorption structures of formic acid on cluster 1, red atoms are oxygen, orange ones copper, white ones hydrogen and grey ones carbon.\newline
	\textbf{A} \, acid hydrogen binds to one \Osuf~ species, doubly bound oxygen coordinates with a \CUS~species.\newline
	\textbf{B} \, carbon-bound hydrogen binds to one \Osuf~species, doubly bound oxygen coordinates with a \CUS~species.\newline
	\textbf{C} \, carbon-bound hydrogen binds to one \Osuf~species, acid oxygen coordinates with a \CUS~species.\newline
	\textbf{D} \, carbon coordinates with an \Osuf~species (top view).}
	\label{fig:ads_structures}
\end{figure}

Adsorption of formic acid will be the first step towards any reaction mechanism on the surface. In order to predict adsorption structures and energies several geometry optimization procedures of formic acid on the \Cu2O (111) surface were performed. This procedure yielded two stable adsorption structures. Two (A and B) exhibit a large adsorption energy (50 to 85 kcal/mol) at both levels of theory and are thus considered to be stable. Both have a bridged structure where either the acid hydrogen or the carbon-bound hydrogen binds to one \Osuf~species and the doubly bound oxygen coordinates with a \CUS~species. Structure C has a high adsorption energy (about 50 kcal/mol) within Turbomole's b-p, SVP level of theory but DMol$^3$'s BP/DNP level of theory predicts only 6 kcal/mol adsorption energy. Structure C is not stable within the b-p/SVP level of theory in Turbomole, a geometry optimization proceeds towards Structure A.  The reason for the qualitative  discrepancy between the two levels of theory in describing structures C and D has yet to be investigated. See Figure \ref{fig:ads_structures} for the adsorption structures on cluster 1, where only parts of the surface are displayed for clarity. Table \ref{tab:ads_energies} lists adsorption energies for the different structures within both theories and on the two Cu$_{70}$O$_{35}$ cluster models. While adsorption energies are generally larger within b-p/SVP; both levels of theory agree on the general trend that the carboxyl-bridged structure is energetically more favourable. Note, that adsorption energy is the negative reaction energy of the adsorption reaction. Thus, a positive adsorption energy indicates a stable adsorption.

The two Cu$_{70}$O$_{35}$ cluster models have been used to find adsorption structures and the results obtained with both agree on structure and the general trend of the adsorption energy of the structures.

\vspace{0.5cm}

\begin{table}
\begin{center}
\begin{tabularx}{0.8\textwidth}{|l|X|X|X|X|} \hline
 & DMol$^3$, BP, DNP basis  & \multicolumn{2}{|c|}{Turbomole, b-p, SVP basis}\\
\hline
 & Cluster 1 & Cluster 1 & Cluster 2\\
\hline
\textbf{Structure A} & 30 & 85 & 65\\
\hline
\textbf{Structure B} & 17 & 68 & 50\\
\hline
\textbf{Structure C} & 6 & 63 & 40\\
\hline
\textbf{Structure D} & 5 & not stable & not stable\\
\hline
\end{tabularx}
\end{center}
\caption{adsorption energies in kcal mol$^{-1}$.}
\label{tab:ads_energies}
\end{table}

\clearpage

\section{Decomposition and Reaction Paths}
\subsection{Vibrational Analysis of the adsorbed Formic Acid Molecule}

A vibrational analysis of the adsorbed formic acid molecule has been carried out. Most importantly, vibrational data is a link to experimental work as it allows experimentalists to search for surface species when theoretical spectra are available and it allows to verify the theoretical model when spectra of known surface species are available. The spectra of the two adsorbed species allow to distinguish between them in an experimental spectrum. The normal modes from such calculations are also starting points for transition state searches in the eigenvector following scheme, as explained in \ref{sect:EV_following}.

\begin{figure}[!hb]
\includegraphics[width=14.48cm]{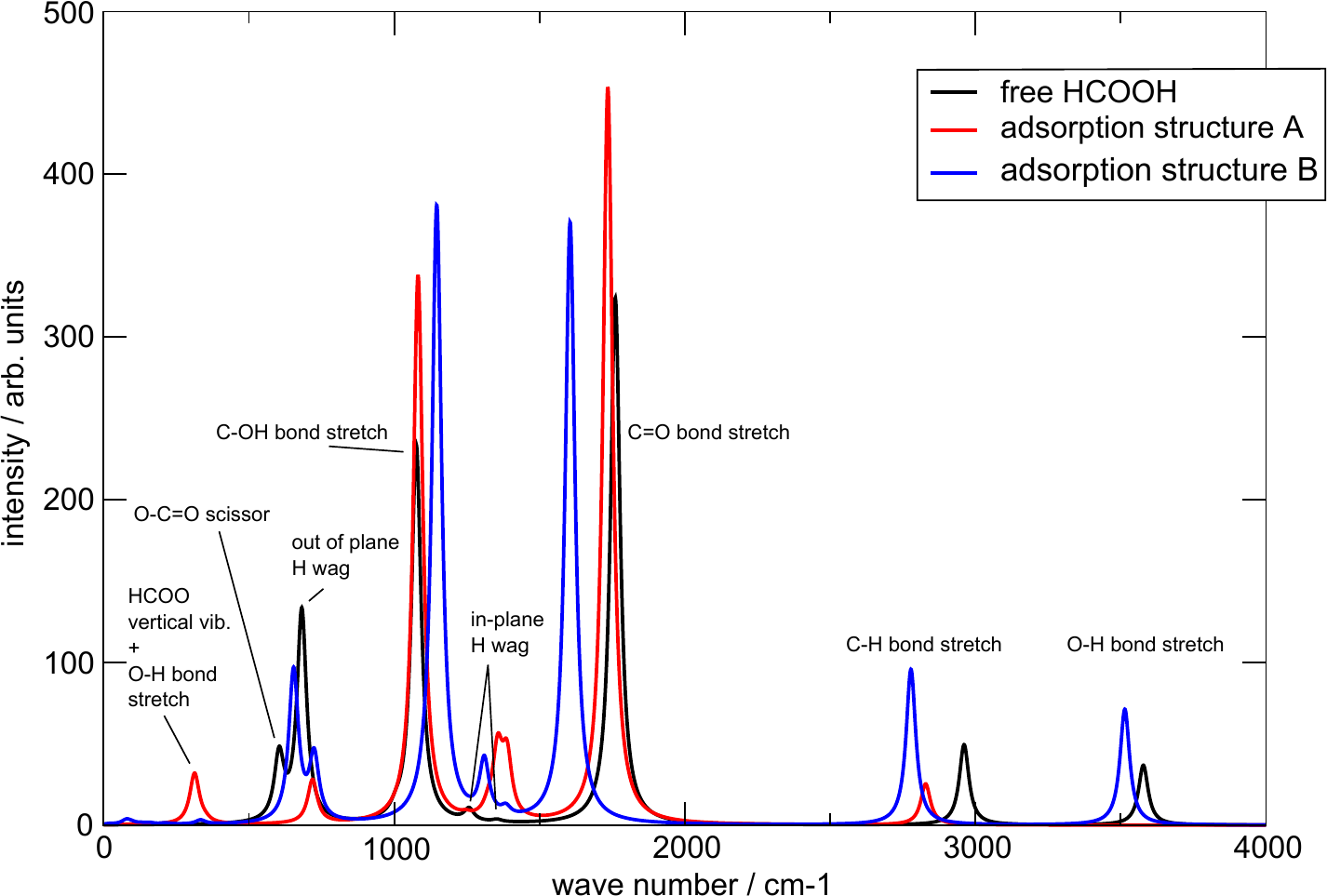}
		\caption{Infrared (vibrational) spectra of formic acid, comparing the free molecule with the adsorbed species.}
		\label{fig:spectra}
\end{figure}

Figure \ref{fig:spectra} shows the vibrational spectrum of formic acid for comparison of the free molecule with the adsorbed structures A and B. For the adsorbed species, only vibrations of the adsorbate atoms were calculated, freezing the surface cluster's atoms. Mostly, the frequencies change only little, but the trend is to red-shift (soften) for bond-stretching modes and to blue-shift (harden) for H-wagging modes.
The O-H stretching vibration of the first adsorption structure disappears at 3578 cm$^{-1}$ but a mode at 314 cm$^{-1}$ appears. It is still a stretch of the O-H bond but when the formic acid molecule is adsorbed, the acid proton binds to the surface and the formate rest vibrates vertically in this mode. Thus, the mass of the vibrating element (49~u) is much larger than that of a single proton in the O-H stretch of the free formic acid. The acid proton is likely in a soft double well potential of the two neighbouring oxygen atoms, which would account for a great part of that effect.

Additional modes in the range from 40 to 275 cm$^{-1}$ emerge when the molecule adsorbs, but these are translations and rotations of the molecule on the surface and do not lead to configurations relevant for reactions.

\subsection{Reaction Modelling using Linear Synchronous Transit}

\begin{figure}
\begin{overpic}[width=\textwidth]{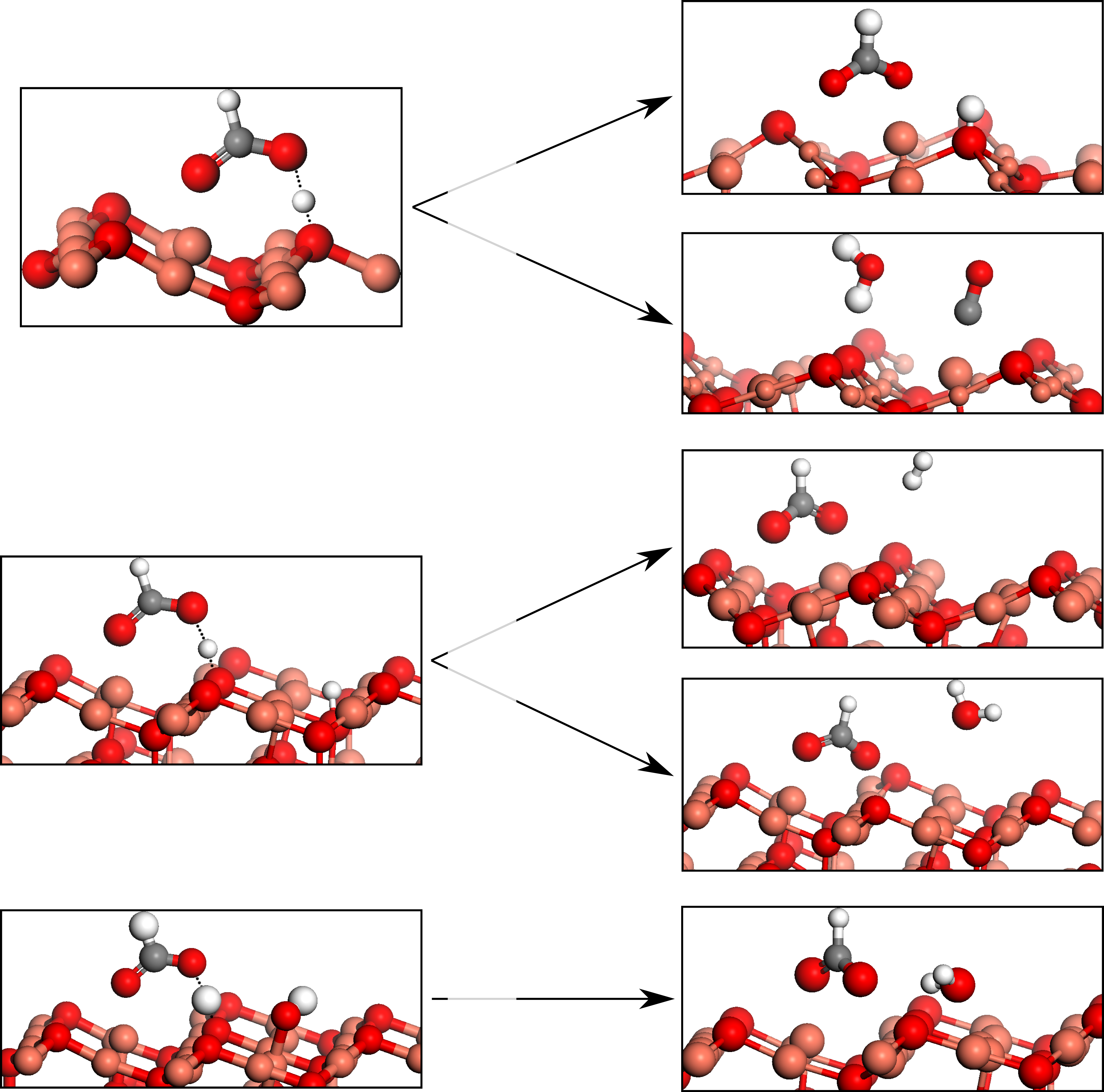}
	\put(50,92){${\scriptstyle \Delta E\,=\,7 }$}
	\put(50,90){$\scriptstyle E_\text{a}\,=\,10 $}
	\put(50,88){$\scriptstyle k\,=\,10^7 $}
	\put(41.5,81){$ \ref{rctn:HCOOH_acid_decomp} $}
	
	\put(50,68){$\scriptstyle \Delta E\,=\,-16 $}
	\put(50,66){$\scriptstyle E_\text{a}\,=\,80 $}
	\put(50,64){$\scriptstyle k\,=\,10^{-31} $}
	\put(41.5,76){$ \ref{rctn:HCOOH_CO_decomp} $}
	
	\put(50,52){$\scriptstyle \Delta E\,=\,-35 $}
	\put(50,50){$\scriptstyle E_\text{a}\,=\,84 $}
	\put(50,48){$\scriptstyle k\,=\,10^{-33} $}
	\put(41,40){$ \ref{rctn:acidH+adsH} $}
	
	\put(50,28){$\scriptstyle \Delta E\,=\,-18 $}
	\put(50,26){$\scriptstyle E_\text{a}\,=\,47 $}
	\put(50,24){$\scriptstyle k\,=\,10^{-14} $}
	\put(41,36){$ \ref{rctn:OSUF+adsH} $}
	
	\put(50,9){$\scriptstyle \Delta E\,=\,-3 $}	
	\put(41,7){$ \ref{rctn:acidH+adsOH} $}
	
\end{overpic}
\caption{Overview of the modelled reactions, with reaction energies and barriers in kcal mol$^{-1}$ and reaction rate constants in s$^{-1}$}
\label{fig:rctn_overview}
\end{figure}
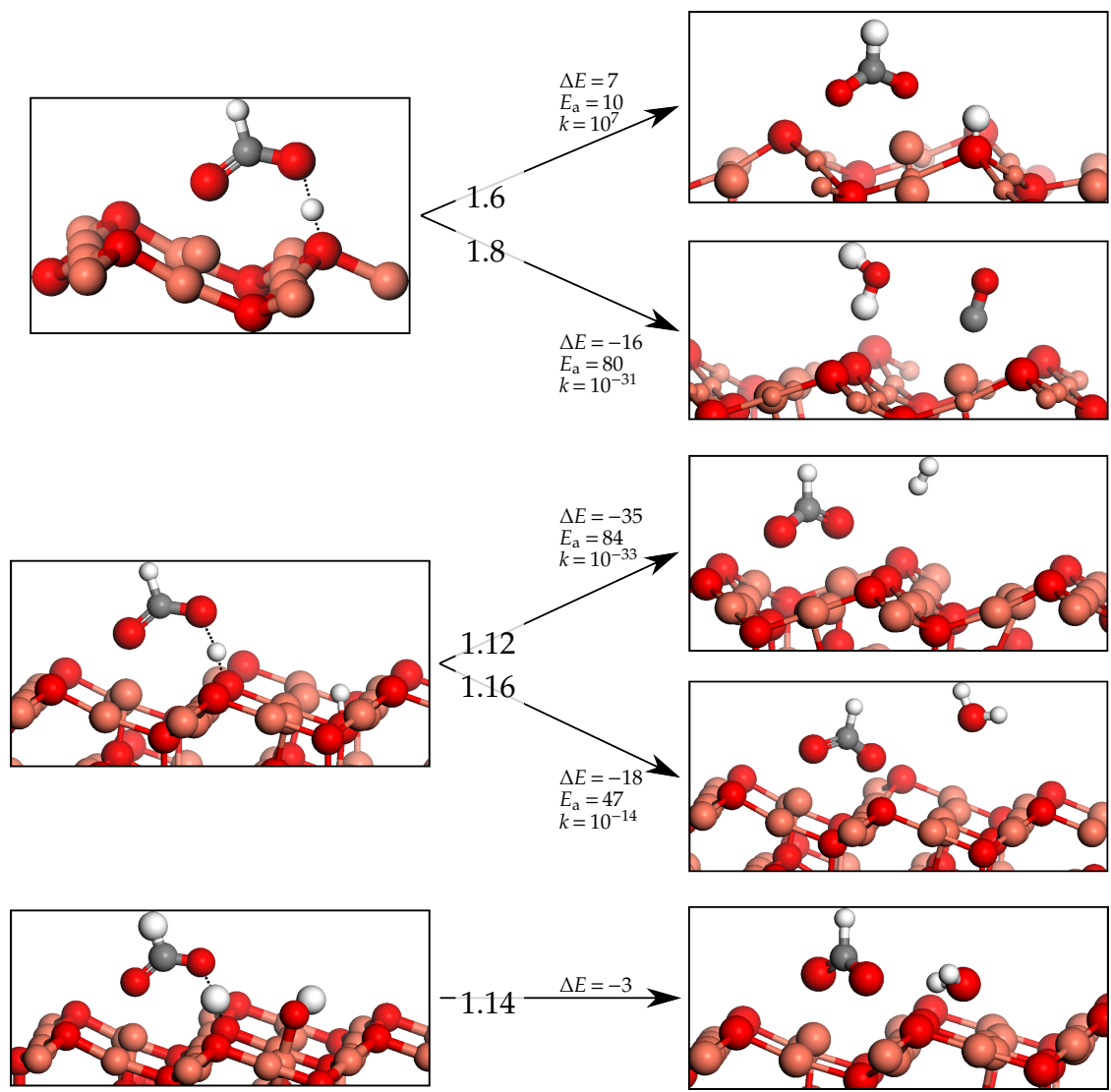

In order to evaluate the model's applicability for the prediction of reaction mechanisms and energies and to gain a first understanding about possible reactions on the surface, some of the reactions from section \ref{chap:reactions} were modelled in a synchronous transit scheme as implemented the DMol$^3$ program. It must be noted that the results are upper bounds to the reaction barriers but a general trend on kinetic probability of each reaction will be discernible.
Two of the suspected decomposition reactions (\ref{rctn:HCOOH_acid_decomp} and \ref{rctn:HCOOH_CO_decomp}) were modelled.
Furthermore, in a technical application the copper oxide surface will never be clean. Adsorbed hydrogen, hydroxyl and water species from wet oxidation pulses as well as organic dissociation products of the copper precursor are likely to saturate the surface.
Thus, some of the reactions with previously adsorbed species have been modelled too.
These are reactions \ref{rctn:acidH+adsH}, \ref{rctn:acidH+adsOH} and \ref{rctn:OSUF+adsH}.
Note that this is not intended as an exhaustive search for a reaction mechanism, but as a starting point for ongoing research and to gain an initial understanding on how reactions may be modelled with the surface structure at hand.
Table \ref{tab:LST_reactions} lists reaction energies and barriers as predicted by an LST+QST search and estimated rate constants $k$ (see below).
All modelled reactions are illustrated in figure \ref{fig:rctn_overview}.

\begin{table}[!hb]
\begin{center}
\begin{tabular}{|c|c|c|c|c|}
\hline \# & reaction & $\Delta E$ & $E_\text{a}$ & $k$ \\ 
\hline \ref{rctn:HCOOH_acid_decomp} & \cee{HCOOH_{(ads)} -> H_{(ads)} + HCOO_{(ads)}} & 7 & 10 & \raisebox{0pt}[3pt]{$10^{7}$} \\ 
\hline \ref{rctn:HCOOH_CO_decomp} &  \cee{HCOOH_{(ads)} -> CO_{(ads)} + H2O_{(ads)}} & -16 & 80 & $10^{-31}$ \\ 
\hline \ref{rctn:acidH+adsH} & \cee{HCOOH_{(ads)} + H_{(ads)} -> HCOO_{(ads)} + H_{2(g)}} & -35 & 84 & \raisebox{0pt}[3pt]{$10^{-33}$} \\ 
\hline \ref{rctn:acidH+adsOH} & \cee{HCOOH_{(ads)} + OH_{(ads)} -> HCOO_{(ads)} + H2O_{(ads)}} & -3 & n.a. & n.a. \\ 
\hline \ref{rctn:OSUF+adsH} & \cee{O_{SUF}-HOOCH +  H_{(ads)} -> HCOO_{(ads)} + H2O_{(g)}} & -18 & 47 & \raisebox{0pt}[3pt]{$10^{-14}$} \\ 
\hline 
\end{tabular}
\end{center}
\caption{reaction energies and barriers in kcal mol$^{-1}$ and rate constants at 400K in s$^{-1}$.}
\label{tab:LST_reactions}
\end{table}

Most of the reactions are exothermic, but the pure acid deprotonation to yield an adsorbed proton is slightly endothermic.
In contrast to the gas phase reaction, the decomposition to CO and \cee{H2O} is exothermic but still has a high activation barrier of $\pm 80\,\text{kcal mol}^{-1}$.
Both the acid deprotonation to form molecular hydrogen with an adsorbed proton and to form water with surface oxygen and an adsorbed proton are strongly exothermic but hindered by high reaction barriers. Reaction \ref{rctn:OSUF+adsH} is an actual reduction reaction, in the sense that it removes an oxygen atom from the surface. A reaction with adsorbed OH to form water is slightly exothermic, the reaction barrier has yet to be calculated for this reaction.

An accurate prediction of reaction kinetics from first principles is difficult. It is not possible to report actual error bars on any electronic structure calculation, and even when assuming a very small error in the reaction barrier of 1 kcal/mol the reaction rate constant has an uncertainty of one order of magnitude because the activation energy enters an exponential function. However, from the available data, a rough estimate of reaction rates can be made according to the \textsc{Eyring}-Equation (\ref{eqn:Eyring}), assuming zero for $ \Delta S $ and $ \Delta H_\text{vib} $. These are expected to introduce less than $\pm 5\,\text{kcal mol}^{-1}$ in the exponent from the findings in \cite{my_Schmeisser2011}.

It can be predicted, that the pure deprotonation by reaction \ref{rctn:HCOOH_acid_decomp} will not be limited from the intrinsic reaction rate but rather from the amount of free active sites on the surface. In contrast, the latter two deprotonation reactions are certainly kinetically unfavourable because of their high activation barrier. Reaction \ref{rctn:OSUF+adsH} might proceed slowly, a rough estimate for the reaction rate at 400\,K is $10^{-14}~\text{s}^{-1}$. But reaction \ref{rctn:acidH+adsH} has a far too high barrier and will not proceed at 400\,K before other reactions take place. A reaction rate constant is estimated to about $10^{-33}~\text{s}^{-1}$.

To present a more comprehensive quantity, the theoretical half-life of a surface population with one of the reactants that undergoes only one reaction is given by $t_{\frac{1}{2}} = \ln{2} / k$. The half-life times for the three high-barrier reactions are $10^{30}~\text{s}$, $10^{33}~\text{s}$ and $10^{12}~\text{s}$, so even if the reaction barriers are overestimated by $10\,\text{kcal mol}^{-1}$, no significant conversion will take place through these reactions.

The results show, that a kinetic evaluation of model reactions within the limitations of the applied theory is possible with the model system at hand and can be used to test for probable and improbable reactions and mechanisms.

Concerning a reduction mechanism for the copper oxide layer, it is clear that other mechanisms with lower kinetic hindering must be existent or catalytic influence will be necessary in a technical application.

All reactions were modelled using uncharged structures. Mulliken partial charges of some of the adsorbed species are listed here. After reaction \ref{rctn:HCOOH_acid_decomp}, the adsorbed hydrogen has a charge of $0.286\,$e and the formate rest has a total charge of $-0.472\,$e, where e is the elementary charge. The adsorbed single hydrogen in the reactant structure for reactions \ref{rctn:acidH+adsH} and \ref{rctn:OSUF+adsH} has a charge of $0.171\,$e. In the reactant structure of reaction \ref{rctn:acidH+adsOH}, the hydroxyl oxygen has a charge of $-0.566\,$e and the hydrogen has a formal charge of $0.264\,$e. A thorough discussion of the bonding mechanisms and electron density distributions should be done in future projects but would go beyond the scope of this work.

\subsection{Transition State Searches using Eigenvector Following}

Line scans of the vibrational modes of adsorption structure 1 have been performed. 

A complete assessment of the chemically interesting modes was not possible in the time frame of this work.
One transition state between two adsorption sites could be localized. However, no converged structure of transition states for decomposition reactions can be reported from eigenvector following calculations of the negatively curved line scans.

The results show, that a full evaluation of reaction barriers from an eigenvector following scheme will take significant effort that would go beyond the scope of this work. Eigenvector following methods might however prove useful for further validation and refinement of the results obtained from synchronous transit methods.

\begin{figure}[!hb]
\begin{center}
	\includegraphics[width=0.7\textwidth]{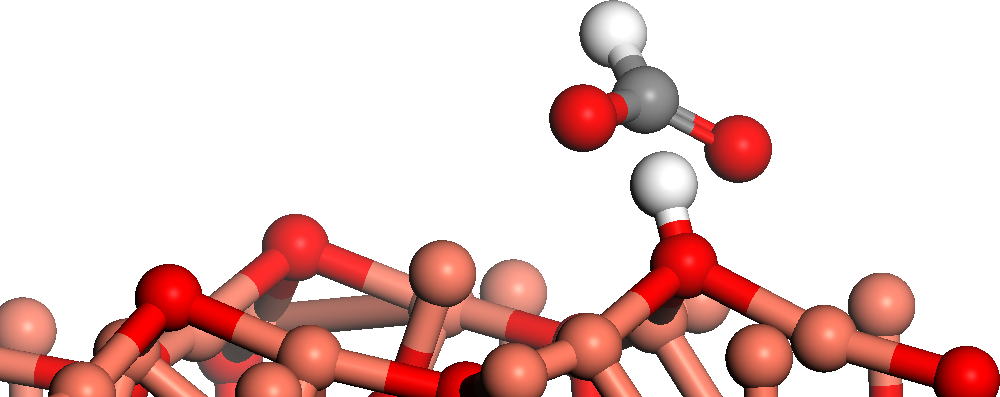}
	\caption{Transition state for mobility of the adsorbate from one \CUS~to another.}
	\label{fig:mobilityTS}
\end{center}
\end{figure}

Figure \ref{fig:mobilityTS} shows the transition state for mobility of the adsorbate from one \CUS~to another. The oxygen atoms in the formic acid will change roles during this reaction. The reaction barrier is 24 kcal mol$^{-1}$.

\chapter{Summary and Outlook}

Four cluster models for a copper(I)oxide (111) surface have been designed, of which three were studied with respect to their applicability in density functional calculations in the general gradient approximation. The two smallest systems have been found to be computationally feasible and produce qualitatively matching results, investigation of the other structures is an ongoing project. Further validation will be necessary for quantitative results. The larger clusters are similar in design to the smaller ones and may by comparison provide information about border effects and convergence with system size. The models exhibit the active sites of the \mbox{\Cu2O(111)} surface and are capable of predictions about adsorption and subsequent reactions. Reaction energies and barriers can be estimated in the scope of density functional theory calculations. Further validation against a slab model with periodic boundary conditions and experimental data should be performed in further work.

Formic acid adsorption on these systems was modelled and yielded four different adsorption structures, of which two were found to have a high (50-85 kcal/mol) adsorption energy. Vibrational spectra of adsorption structures A and B are available. These are important for experimental distinction of the two structures and allow further validation of the calculations once experimental data is available. 

The energetically most favourable adsorption structure (A) was further investigated with respect to its decomposition and a few reactions with adsorbed H and OH species using synchronous transit methods to estimate reaction barriers and single point energy calculations for the reaction energy. 
First, the acid deprotonation of formic acid on the surface was modelled.
 The reaction is reported to be endothermic ($ \Delta E = 7 \, \text{kcal mol}^{-1} $) with a modest reaction barrier of $ 10 \,  \text{kcal mol}^{-1} $. It is thus not limited by reaction kinetics but by formic acid concentration and adsorption sites. 
The decomposition to CO and \cee{H2O} was found to be exothermic ($ \Delta E = -16\,\text{kcal mol}^{-1} $) but is inhibited by a high barrier of $84\,\text{kcal mol}^{-1}$. 
The reaction with an adsorbed proton to form molecular hydrogen causes an exothermic energy change of $-35\,\text{kcal mol}^{-1}$ but it also has a high barrier of $84\,\text{kcal mol}^{-1}$. 
One actual reduction reaction that would remove a surface oxygen to form water with the acid hydrogen and another adsorbed proton is also exothermic ($\Delta E = -18\,\text{kcal mol}^{-1}$) but hindered by a barrier of $47\,\text{kcal mol}^{-1}$. 

Although reaction barriers, as calculated so far, are upper bounds and rate constants are strongly dependant on the barrier, no change in the qualitative trend is expected because the barriers are very high. Thus, a reaction mechanism for the reduction of copper oxide by formic acid must go through different reactions not considered here or needs catalytic support in the surface, for example by ruthenium inclusion as investigated in \cite{my_MuellerSCD2011}.

The results show, that the presented cluster models can be employed for electronic structure calculations at the density functional level to compute reaction energies and barriers. All \mbox{\Cu2O(111)} surface reconstructions discussed in literature can easily be modelled using the existing structures.

Thus, the present work may serve as a basis for detailed studies of reactions on the \Cu2O (111) surface which might include further reactions involving different surface species, more spectroscopic data for comparison with experimental work and a thorough investigation of the chemical bonding mechanisms. If necessary more accurate electronic structure methods than DFT might be employed.

\chapter*{Acknowledgement}

\begin{center}
My thank goes to the Fraunhofer ENAS institute and it's director Prof. Dr. Thomas Gessner as well as  Prof. Dr. Stefan Schulz for making this work possible and providing the necessary resources.

\vspace{0.3cm}

Special thank goes to Dr. Jörg Schuster for his trust, his continued assistance, friendly mentoring and for all the countless discussions.

\vspace{0.3cm}

I greatly appreciate Prof. Dr. Alexander Auer's priceless telephone lectures on quantum chemistry, his continued effort in correcting my work and making the visits to the Max-Planck-Institute for Iron Research possible.

\vspace{0.3cm}

I would also like to thank Dr. Ulrich Biedermann for his immense help in modelling the cluster structures and Udo Benedikt for his roadside assistance whenever the Turbomole vehicle wouldn't run.

\vspace{0.3cm}

Thomas Wächtler and Steve Müller have both been very helpful and supportive in numerous discussions.

\vspace{0.3cm}

I am very grateful to my colleagues in the simulation lab for lots of good discussions, for putting up and watering plants and for such a nice working atmosphere as well as to the morning coffee group for sharing relaxing breaks, coffee and the latest gossip.

\vspace{0.3cm}

Also, the \textbf{C}hemnitz \textbf{H}igh-Performance-L\textbf{i}nux-\textbf{C}luster (CHiC) project and the people who make it work deserve praise.

\vspace{0.3cm}

It is important to me to also mention all the busy developers of the numerous programs I used every day : molden \cite{molden}, Inkscape, The GIMP, pdf-\LaTeX, bibLaTeX, TeXmaker, Mendeley Desktop, Xming and PuTTY.

\vspace{0.3cm}

Additional thank goes to my family and friends and, especially, to Alice who all were very supporting and understanding.
\end{center}

\printbibliography

\end{document}